\documentclass{ar-1col-S2O}
\usepackage[numbers]{natbib}
\usepackage{url}
\usepackage{bm}
\usepackage[hidelinks]{hyperref} 

\jname{Xxxx. Xxx. Xxx. Xxx.}
\jvol{AA}
\jyear{YYYY}
\doi{10.1146/annurev-nucl-102622-011319}

\begin{document}

\markboth{M. Casarsa, D. Lucchesi, L. Sestini}{Experimentation at a muon collider}

\title{Experimentation at a muon collider}

\author{Massimo Casarsa,$^1$ Donatella Lucchesi$^2$ and Lorenzo Sestini$^3$
\affil{$^1$INFN Sezione di Trieste, via A. Valerio 2, I-34127 Trieste, Italy; email: massimo.casarsa@ts.infn.it}
\affil{$^2$ Dipartimento di Fisica e Astronomia, Universit\'a di Padova and INFN Sezione di Padova Via Marzolo 8, 35131 Padova, Italy; email: donatella.lucchesi@pd.infn.it}
\affil{$^3$INFN Sezione di Padova, via F. Marzolo 8, I-35131 Padova, Italy; email: lorenzo.sestini@pd.infn.it}}

\begin{abstract}
Experimental activities involving multi-TeV muon collisions are a relatively recent endeavor. The community has limited experience in designing detectors for lepton interactions at center-of-mass energies of 10 TeV and beyond. This review provides a short overview of the machine characteristics and outlines potential sources of beam-induced background that could impact the detector performance.

The strategy for mitigating the effects of beam-induced background on the detector at $\sqrt{s}=3$ TeV is discussed, focusing on the machine-detector interface, detector design, and the implementation of reconstruction algorithms.

The physics potential at this center-of-mass energy is evaluated using a detailed detector simulation that incorporates the effects of beam-induced background. This evaluation concerns the Higgs boson couplings and the Higgs field potential sensitivity, that then are used to get confidence on the expectations at 10 TeV.

The physics and detector requirements for an experiment at $\sqrt{s}=10$ TeV, outlined here, form the foundation for the initial detector concept at that center-of-mass energy.
\end{abstract}

\maketitle

\tableofcontents


\section{INTRODUCTION: YET ANOTHER COLLIDER?}
Muon colliders have a long history. The first mention of a muon storage ring in the literature dates back to 1965~\cite{mu_rings}.
Several developments were carried out in the following years, and the proposal of a muon collider appeared on the US physics scene at Snowmass 1996~\cite{SM96}. A systematic study of a multi-TeV muon collider started as a project in the US in 2011 with the Muon Accelerator Program (MAP)~\cite{mark, MAP} to investigate the feasibility of a facility where both neutrino physics and experiments based on muon collisions could be possible.
MAP designed colliders with center-of-mass energies of 1.5, 3, and 6 TeV\footnote{In this paper natural units where $\hbar=c=1$ are used.} and optimized the interaction region (IR) configuration for a collider operating at $\sqrt{s}=1.5$ TeV. 
After the recommendations of \textit{The 2020 Update of the European Strategy for Particle Physics}~\cite{ESPP}, an International Muon Collider Collaboration (IMCC)~\cite{imcc} was formed 
at CERN with the aim of establishing a feasible design and defining the necessary R\&D for a multi-TeV muon collider. The Snowmass 2021 process~\cite{snowmass2021}, a community planning exercise in particle physics, has strengthened the collaboration between the US and Europe, 
allowing for a revision of the studies performed by MAP and progress in the design of the facility. The current focus is on the definition of the parameters of a muon collider at $\sqrt{s}=3$ and $\sqrt{s}=10$ TeV, but the project still foresees a facility that can also host neutrino experiments.
%
%
\subsection{Muon collider concept}
\label{sec:intro-machine}

\begin{figure}[t]
\centering
\includegraphics[width=0.9\textwidth]{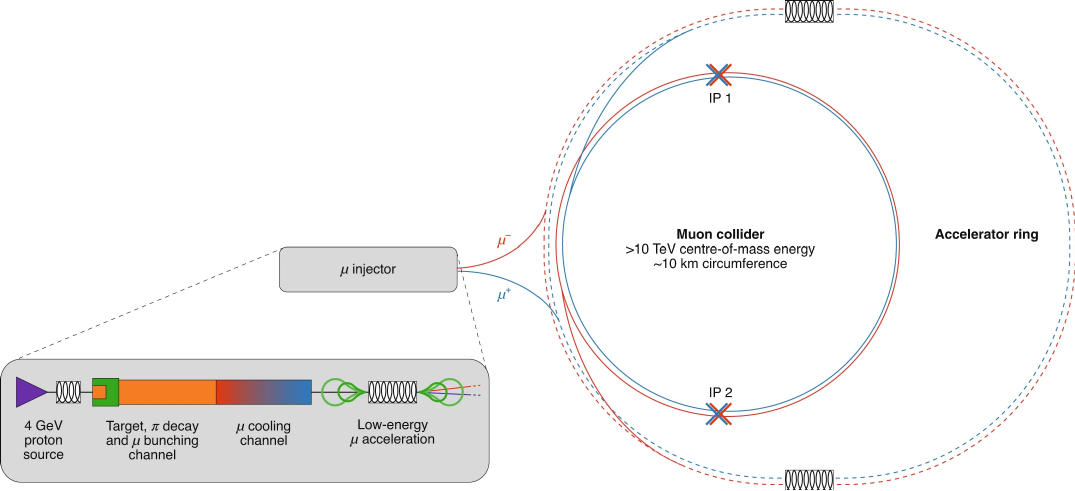}
\caption{Conceptual layout of a multi-TeV muon collider.
\label{fig:facility}}
\end{figure}

The design of a muon collider facility is mainly driven by the short lifetime of muons, $2.2~\mu$s at rest. A conceptual layout of such a facility is presented in \textbf{Figure~\ref{fig:facility}}. It foresees a muon injector and two rings, a large one for muon acceleration and a small one for collisions. The reduced dimensions of the collider are required to allow muons to pass through the interaction point (IP) many times before decaying. The current ring sizes~\cite{muc_epjc} do not take into account any specific site, and by adjusting the machine configuration, they can potentially exploit existing and planned infrastructures at CERN and Fermilab.
Although in principle the collider could operate with two interaction points, in this paper only one interaction point and one detector are assumed.
Preliminary parameters of the muon collider complex~\cite{muc_epjc}, relevant for the following, are summarized in \textbf{Table~\ref{tab:parameters}}. It must be noticed that the instantaneous luminosity increases significantly from $\sqrt{s}=3$ TeV to $\sqrt{s}=10$ TeV. As demonstrated in~\cite{muc_epjc}, it grows approximately with the square of the muon beam energy at fixed average bending magnetic field. This is a feature of the muon collider due to the fact that the beam can be recirculated multiple times through the interaction point thanks to the negligible effect of the  beamstrahlung on the focusing achievable at the interaction point. The luminosity also benefits from having few high-intensity bunches per beam: the MAP design found the optimum with one bunch of $\mu^+$ and one bunch of $\mu^-$ at a time in the collider ring~\cite{singleB}. 
\begin{table}[b]
\tabcolsep7.5pt
\caption{Preliminary muon collider parameters based on a revised MAP design. }
\label{tab:parameters}
\begin{center}
\begin{tabular}{@{}llcc@{}}
\hline
Parameter & Unit of measure & \multicolumn{2}{c}{Target value}\\
\hline
Center-of-mass energy ($\sqrt{s}$)  & TeV  & 3 & 10\\
Luminosity  &$10^{34}$ cm$^{-2}$ s$^{-1}$ &2 & 20 \\
Collider circumference  &km &4.5 & 10 \\
Muons/bunch   &$10^{12}$ &2.2 &1.8\\
Beta function at IP & mm& 5 & 1.5\\
\hline
\end{tabular}
\end{center}
\end{table}

Muon decays along the collider ring produce intense fluxes of high-energy neutrinos that can generate hadronic and electromagnetic showers when interacting with the underground environment.
In particular, intense and highly collimated fluxes are produced in the straight sections of the machine. The effects at the Earth's surface, where the neutrinos emerge, even hundreds of kilometers away from the collider complex, are under study to ensure that the impact on the environment remains significantly below the legal limits and those of the LHC.  The complete determination of such effects is a multi-parameter problem: they depend on the machine center-of-mass energy, the collider configuration and the geological characteristics of the soil. 

The heart of the facility is the muon injector. Intense proton beams, produced by a proton source, hit a multi-MegaWatt class target to generate pions that eventually decay to muons. The target is immersed in a high solenoidal magnetic field to capture and guide pions into the decay channel, where muons are captured in a bunch train, with a time dependent acceleration due to their different energies. 

Muon cooling process is critical to achieve the necessary beta function at IP (see \textbf{Table~\ref{tab:parameters}}) for the high luminosity at any center-of-mass energy. The challenge of the cooling process is the muon short lifetime, the cooling must take place more quickly than any of the cooling methods presently in use.
A new technology, called “ionization cooling”, has been tested by the Muon Ionization Cooling Experiment (MICE)~\cite{MICE}. The particle momentum is reduced through ionization energy loss in absorbers and replenished only in the beam direction through radio-frequency cavities. MICE experimentally demonstrated the first part, it measured an increase in the phase-space density of the beam, and the results agree with the simulations. The cooling and the re-acceleration of the particles in the direction of the beam have to be proven, although well predicted by the simulation tuned on MICE experimental results.

The first step toward a muon collider facility is the demonstration of the 6D cooling at low emittance and the re-acceleration of the muon beam through several cooling cells \cite{6dcooling}. This requires, as highlighted in~\cite{muc_epjc}:
\begin{itemize}
    \item[-] design, construction and integration of the cooling cell;
    \item[-] a sufficiently intense proton beam that impinges on a target to produce pions, typically with a momentum range of 100–300 MeV, which in turn decay into muons with large transverse emittance;
    \item[-] an upstream instrumentation system to measure the muon beam properties and then guide it to a certain number of cooling cells;
    \item[-] a downstream instrumentation apparatus that measures the beam characteristics.
\end{itemize}
The design of this experimental demonstration with the relative instrumentation is part of the current activities of the IMCC. This muon beam facility could be hosted in any laboratory where an adequate proton source is available.
Such a facility could also host physics experiments such as the muon beam dump project~\cite{cari}. The proposed study of new gauge forces would require a beam with energy of the order of 100 GeV, in addition to the dedicated detector. The feasibility of this experiment will be assessed during the design phase of the demonstration facility.
%
\subsection{The beauty of colliding muons}
\label{sec:intro-physics}
Muon collisions at high center-of-mass energies open a completely new physics scenario. A muon collider offers the possibility to reach high center-of-mass energy with high luminosity and also to perform high precision measurements thanks to the fact that muons are elementary particles. 

The direct particle production can proceed via $s$-channel $\mu^+\mu^-$ annihilation, in which heavy particles of mass $M$
can be produced with a kinematical threshold at $M=\sqrt{s}/2$. As pointed out in~\cite{Tao}, states such as $Z^{\prime}$, $W^{\prime}$, heavy Higgses or other heavy particles could be produced singly in association with soft/collinear vector bosons like $\gamma$, $Z^0$ and $W^{\pm}$ overcoming the kinematical limit. Comparisons of the production cross sections at proton and muon colliders for $2\rightarrow 2$ and $2\rightarrow 1$ processes are studied in~\cite{muc_forum_report, maltoni, ruiz}, 
taking into account the differences between electroweak and colored states, in order to gain insight into the most appropriate collider for exploring the very high energy frontier. Such a comparison is not trivial, because it cannot represent the peculiarities of the two machines and the complexity of the data produced in the two very different environments.

Multi-TeV muon beams colliding at energies much higher than the electroweak scale, $\sim$100 GeV, have a high probability of emitting electroweak bosons. A summary of the theoretical investigations carried out so far on the characteristics of such an emission is provided in~\cite{muc_epjc}, where it is also mentioned that the experimental observation of these phenomena and their comparison with the theoretical predictions is in itself a very interesting measurement. At the center-of-mass energies considered, the vector boson fusion (VBF) is the dominant production process and this is why the muon collider is often referred to as a ``vector boson collider''. The production cross sections of Standard Model (SM) processes via VBF are very high, see for example the Higgs cross sections reported in \textbf{Section~\ref{sec:higgs}}, which, combined with the high luminosity of the collider and the very low physics background contributions, allow precise SM tests in a completely unexplored energy regime. Precise knowledge of the SM processes at a multi-TeV energy scale gives the possibility to probe new physics indirectly, extending the reach of direct interactions.
%
%
\section{THE MUON COLLIDER SEEN FROM THE EXPERIMENTAL POINT OF VIEW}
\label{sec:mdi}

\begin{figure}[t]
    \centering
    \includegraphics[width=0.9\textwidth]{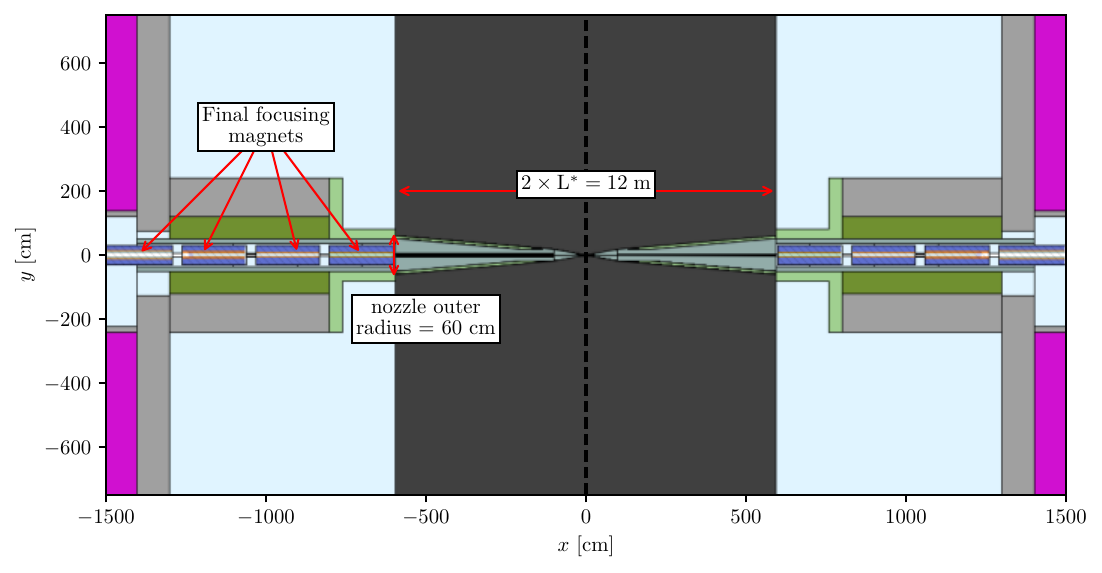}
    \caption{Top view of the IR, the $x$ axis represents the beam direction. The outer shape of the detector is a cylinder with a radius of 6 m. The space between the outer shape and the nozzles is considered as a perfect particle absorber in the beam-induced background simulations. 
    \label{fig:MDI}}
\end{figure}
The design of a detector capable of operating and taking data at a multi-TeV muon collider must take into account not only the physics requirements, but also the IR conditions. 

The studies on the machine-detector interface (MDI) conducted by MAP and more recently by the IMCC indicate that the final focusing magnets cannot be located more than $\pm$6 meters from the IP in order not to degrade the instantaneous luminosity.
\textbf{Figure~\ref{fig:MDI}} shows the configuration of a possible IR, where the available space for the detector along the beam axis is a total of 12 meters. The transverse dimension can be larger, but this would not improve the physics acceptance, especially at very high center-of-mass energy where several relevant physics processes occur forward respect to the beamline, as discussed in \textbf{Section~\ref{sec:physics-10TeV}}.
%
%
\subsection{Background contributions in the detector region}
Elementary particle collisions typically do not have secondary interactions beyond the primary one, as seen in hadron-hadron collisions. However, the muon collider suffers from background coming from the machine, the main sources of which, expected to contribute in the detector region at $\sqrt{s}=3$ and 10 TeV, are:

\begin{description}
\item{\textbf{a) Muon decay}}:
intense muon beams with energies of 1.5 and 5 TeV decay, producing positrons and electrons of several TeV, as well as high-energy photons resulting from synchrotron radiation of the energetic electrons and positrons in the collider magnetic fields.
Most of these particles directly reach the detector and can compromise any physics measurements. 
The MAP project proposed to insert two conical-shaped absorbers, referred to as nozzles~\cite{nozzle}, made of tungsten and covered with borated polyethylene,
inside the detector along the beamline~(\textbf{Figure~\ref{fig:MDI}}).
This combination is effective in absorbing high-energy particles and reducing their impact on the detector.
This shielding structure modifies the characteristics of the background arriving on the detector, as discussed in \textbf{Section~\ref{sec:bck}}.
\\ \noindent
\item{\textbf{b) Incoherent $\bm{e^+e^-}$ production}}: 
the production cross section of $e^+e^-$ pairs in high-energy muon collisions has been calculated in~\cite{incoherent} from the original formulation of Landau and Lifshitz~\cite{landau}. Such a background is synchronous with the beam collisions and is mainly produced along the beam direction. The MAP project evaluated the contribution of this source of background for beams of 750 GeV~\cite{MAP-BIB} and found it to be almost negligible. This is primarily because the production cross-section is suppressed at low muon energy, and the detector's magnetic field effectively traps the $e^+e^-$ pairs near the beamline.
As the beam energy increases, the probability of incoherent positron and electron production rises. These particles could potentially contribute to the background on the inner layers of the detector. Therefore, it is essential to conduct a detailed simulation study taking into account the solenoidal magnetic field value, which is expected to be higher at $\sqrt{s}=10$ TeV.
%
\\ \noindent
\item{\textbf{c) Beam halo}}:
beam losses are unavoidable, but at the moment they are not considered among the background sources in the detector region. It is believed that an appropriately designed collimation system positioned upstream of IP can effectively reduce these losses to a manageable level.
\end{description}


\subsection{Beam-induced background characteristics}
\label{sec:bck}

The nozzles inserted around the IP absorb the high-energy electrons and positrons generated by the muon decays, but are responsible for the fluxes of particles that eventually enter the detector.
Electrons, positrons and photons interacting with the tungsten material of the nozzles and the machine IR elements produce electromagnetic showers.
In addition, hadronic particles, mainly neutrons with a minor contribution from charged hadrons, are generated through photonuclear interactions with the IR material, including the nozzles. The borated polyethylene cladding of the nozzles is designed to absorb neutrons, but it also produces high fluxes of photons, which contribute to background signals in the calorimeter system~(\textbf{Section \ref{sec:calo}}).
Background muons are instead produced through the Bethe-Heitler process~\cite{bethe} by energetic photons from electromagnetic showers.

An assessment of the effects of the beam-induced background (BIB) on the detector response involves a laborious procedure.
It must be noticed that the BIB sample used in the detector studies until now was produced by MAP using the MARS15 software~\cite{mars1, mars2} at a center-of-mass energy of 1.5 TeV. The IR and the nozzle dimensions and composition were optimized to minimize the occupancy on the first layers of the tracking system.

The IMCC has setup a software procedure to simulate the beam induced background~\cite{bib-collamati}:
\begin{itemize}
    \item[1)] The IR accelerator lattice and optics are modeled in the FLUKA code ~\cite{fluka} via the LineBuilder~\cite{LB} package, as well as the nozzles. 
    \item[2)] The detector solenoidal magnetic field is included in the IR description.
    \item[3)] A primary beam is simulated starting from a given distance $L_{BIB}$ from the IP; the value of $L_{BIB}$ depends on the center-of-mass energy and is given by the point along the beamline where the cumulative distribution of particles exiting the machine begins to saturate to 100\%. Given the symmetric nature of $\mu^+$/$\mu^-$, only one primary beam is considered.
    \item[4)] Particles exiting the machine, either from the nozzles or any of the IR components, are saved and then provided as input to \textsc{Geant4}~\cite{geant} simulation software to be tracked inside the detector.
\end{itemize}
The optimization path requires to execute all the actions from 1) to 4) with a given IR and nozzle configuration, and evaluate the occupancy in the detector, until an acceptable/optimal value is reached. The procedure has been carried out using the $\sqrt{s}=1.5$ TeV IR: the results, compared with those obtained with the MARS15 simulation code, show a very good agreement~\cite{bib-collamati}, giving robustness to the evaluation of the beam-induced background in the detector.

\begin{figure}[t]
    \centering
    \begin{minipage}{0.7\textwidth}
        \hspace{1.cm}\textbf{\textsf{a}}\vspace{-0.4cm}

        \centering
        \includegraphics[width=1.0\textwidth]{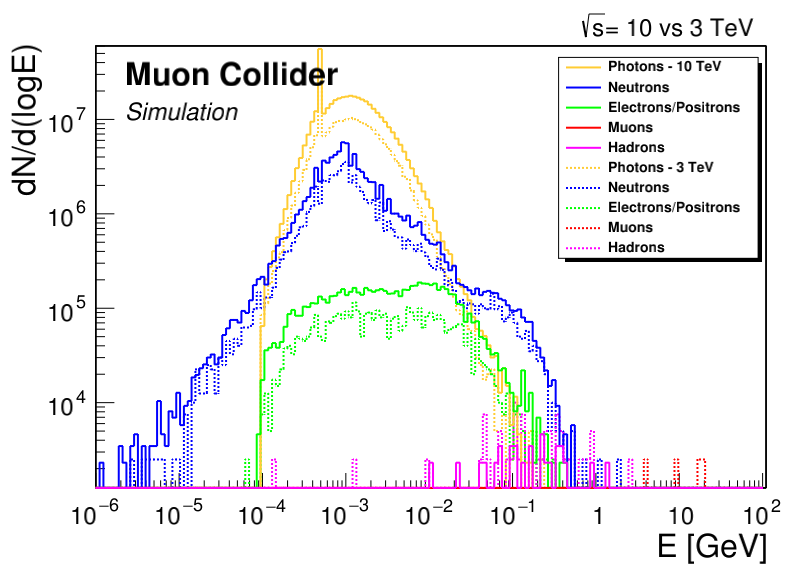}
    \end{minipage} \hfill
    \begin{minipage}{0.7\textwidth}
       \hspace{1.cm}\textbf{\textsf{b}}\vspace{-0.4cm}
       
        \centering
        \includegraphics[width=1.0\textwidth]{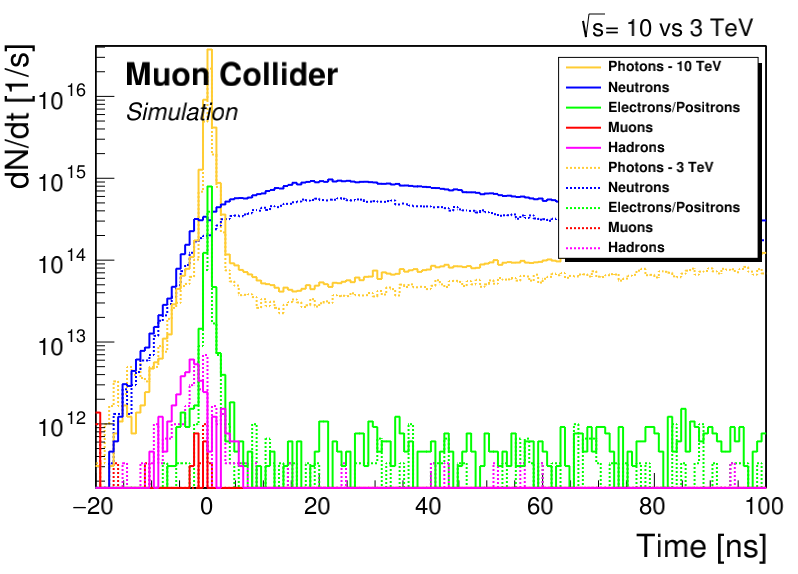}
    \end{minipage}
\caption{\textit{(a)} Energy distribution of particles arriving at the detector for $\sqrt{s}=3$ TeV (dashed line) and $\sqrt{s}=10$ TeV (solid line). Different colors represent different types of particles. 
\textit{(b)} Distribution of the arrival time of the various particle types for the two center-of-mass energies and with the same legend as in the panel on the left.
\label{fig:BIB}}
\end{figure}
The determination of the characteristics of such a background at $\sqrt{s}=3$, 10 TeV is in its early stage. The IR at a center-of-mass energy of 3 TeV~\cite{MAP3TeV} comes from MAP, the preliminary design at $\sqrt{s}=10$ TeV has been done by the IMCC~\cite{IMCC10TeV}. Using the nozzles proposed by MAP~\cite{nozzle}, the beam-induced background has been generated for both machine configurations. The solenoidal magnetic field is set to 3.57 T. 
\textbf{Figure~\ref{fig:BIB}\textit{a}} shows the energy distributions of background particles at $\sqrt{s}=3$ TeV and $\sqrt{s}=10$ TeV for each particle type arriving on the detector in a time window of [-1, 15] ns with respect to the collision time~\cite{bib-daniele, bib-eps}. The two center-of-mass energies show a similar behavior, which is also similar to that at $\sqrt{s}=1.5$ TeV~\cite{muc_epjc}. Given the wide range of energies considered, this demonstrates that the energy distribution of the beam-induced background is determined by the nozzles. \textbf{Figure~\ref{fig:BIB}\textit{b}} illustrates the arrival time of particles at the detector for the same center-of-mass energies. The distributions, similar in the two energy cases, show a significant contribution outside the time window considered adequate for data taking. The similar behavior is expected due to the similar energy spectrum.
It can be noticed that the particle flux at $\sqrt{s}=10$ TeV is slightly higher than that at $\sqrt{s}=3$ TeV, mainly for two reasons: the different center-of-mass energy and  IR configuration.
An important effort is ongoing to optimize the IR and the nozzles shape and materials for the center-of-mass energy of 3 and 10 TeV.

%
%
\section{DETECTOR PERFORMANCE IN THE PRESENCE OF THE BEAM-INDUCED BACKGROUND}
\label{sec:detector}
The design of an experiment at a multi-TeV muon collider has many features in common with the experiments 
at other proposed multi-TeV machines~\cite{ILC,CLIC,FCC,CEPC}, but also presents several novel challenges due to the unique background conditions related to the unstable nature of muons.
The first constraints come from the machine layout and the detector shielding requirements.
As discussed in \textbf{Section~\ref{sec:mdi}}, the available space around the interaction region is limited by 
the collider final focusing magnets and
the MDI, in particular the nozzles reduce the angular acceptance in the forward and backward regions of the detector. 
In addition, each sub-detector 
must deal with the beam-induced background to ensure optimal physics performance.

In the following sections only one detector concept is discussed. Given that the muon collider can have two interaction points, different decisions can be made during the actual design phase.
%
\subsection{Detector requirements and comparison with LHC experiments}
The presence of the BIB makes the detector requirements more similar to those for proton-proton than for positron-electron collisions. In addition, detector technology R\&D is currently focused on the High Luminosity LHC (HL-LHC)~\cite{HL-LHC}, making it interesting to have LHC~\cite{LHC}, with its experiments, and HL-LHC as references. 

The starting point is the comparison of the beam dimensions at the muon collider with the nominal LHC configuration \cite{lhcbeam}: the parameters at the interaction point are reported in 
\textbf{Table \ref{tab:beam}}.
\begin{table}[b]
\tabcolsep7.5pt
\caption{Comparison of the beam size at IP for LHC and muon collider.}
\label{tab:beam}
\begin{center}
\begin{tabular}{@{}l|c|c|c@{}}
\hline
 & LHC & \multicolumn{2}{c}{muon collider} \\
\hline
$\sqrt{s}$ & 14 TeV & 3 TeV & 10 TeV \\
bunch length & 7.7 cm & 5 mm & 1.5 mm \\
transversal bunch size & 16.7 $\mu$m & 3 $\mu$m & 0.9 $\mu$m \\
\hline
\end{tabular}
\end{center}
\end{table}
At the IP, the bunch of the muon collider is shorter and narrower compared to that of the LHC. Consequently, the first detector layer can be positioned closer to the beamline with respect to the ATLAS~\cite{ATLAS} and CMS~\cite{CMS} detectors, to enhance the resolution on the impact parameter and the displaced vertices reconstruction. On the other hand, this is the region where the BIB flux is more intense, and large detector occupancies are expected. Another significant difference with respect to LHC is that the muon collider is expected to operate in single bunch mode and thus the bunch crossing rate coincides with the beam revolution frequency. This results in an expected bunch crossing rate below 100 kHz
(with respect to 40 MHz at LHC) and implies a longer time between two collisions for processing and reading out the detector signals.

Similar to the upgraded ATLAS and CMS experiments for the HL-LHC,
where it will be crucial to precisely measure time to disentangle the pile-up interactions, at the muon collider time measurements are necessary to mitigate the effects of the out-of-time BIB particles. For this purpose, sensors with time resolutions in the 30-100 ps range should be developed. It is clear that the R\&D path towards such precise devices can benefit from synergies with the development of HL-LHC detectors.

Studies on the radiation environment due the BIB at $\sqrt{s}=1.5$ TeV are documented in~\cite{muc_forum_report}. 
\textbf{Table~\ref{tab:radiation}} presents the approximate expected radiation levels per year of data taking and compares them to HL-LHC estimates.
Notably, the maximum total integrated dose is of about 10 Mrad in the region close to the interaction point (distance $R=2.2$ cm), while at HL-LHC it is higher, around 100 Mrad. 
Conversely, the maximum fluence in the calorimeter region ($R=150$ cm) is $\sim$$10^{14}$ 1 MeV neutron equivalent per cm$^2$ at the muon collider, while it is $\sim$$10^{13}$ 1 MeV neutron equivalent per cm$^2$ at the HL-LHC.
Despite some differences in the radiation environment, similar radiation hardness requirements could be expected for the muon collider and the HL-LHC detectors.

At a center-of-mass energy of 10 TeV, the radiation environment is expected to be comparable to that at 1.5 TeV. In fact, both the integrated dose and the maximum fluence are primarily influenced by the BIB, which is comparable in both cases, as discussed in \textbf{Section~\ref{sec:bck}}.

\begin{table}[t]
\tabcolsep7.5pt
\caption{Radiation levels per year at a $\sqrt{s}=1.5$ TeV muon collider and HL-LHC.}
\label{tab:radiation}
\begin{center}
\begin{tabular}{@{}l|c|c@{}}
\hline
 & muon collider & HL-LHC \\
\hline
maximum dose at $R=2.2$ cm & 10 Mrad & 100 Mrad \\
maximum dose at $R=150$ cm & 0.1 Mrad & 0.1 Mrad \\
\hline
maximum fluence at $R=2.2$ cm & $10^{15}$ 1 MeV-neq/cm$^2$ & $10^{15}$ 1 MeV-neq/cm$^2$ \\
maximum fluence at $R=150$ cm & $10^{14}$ 1 MeV-neq/cm$^2$ & $10^{13}$ 1 MeV-neq/cm$^2$ \\
\hline
\end{tabular}
\end{center}
\end{table}
%
%
\subsection{Muon collider detector concept at $\sqrt{s} = 3$ TeV}

The design of the first muon collider detector concept at $\sqrt{s}=3$ TeV was based on CLIC detector~\cite{clic_detector}, which had been optimized for $e^+e^-$ collisions up to 3 TeV.
However, modifications were necessary to adapt it to the muon collider MDI, described in \textbf{Section~\ref{sec:mdi}}, and to mitigate the effects of the BIB. 

\begin{figure}[t]
\centering
\includegraphics[width=0.6  \textwidth]{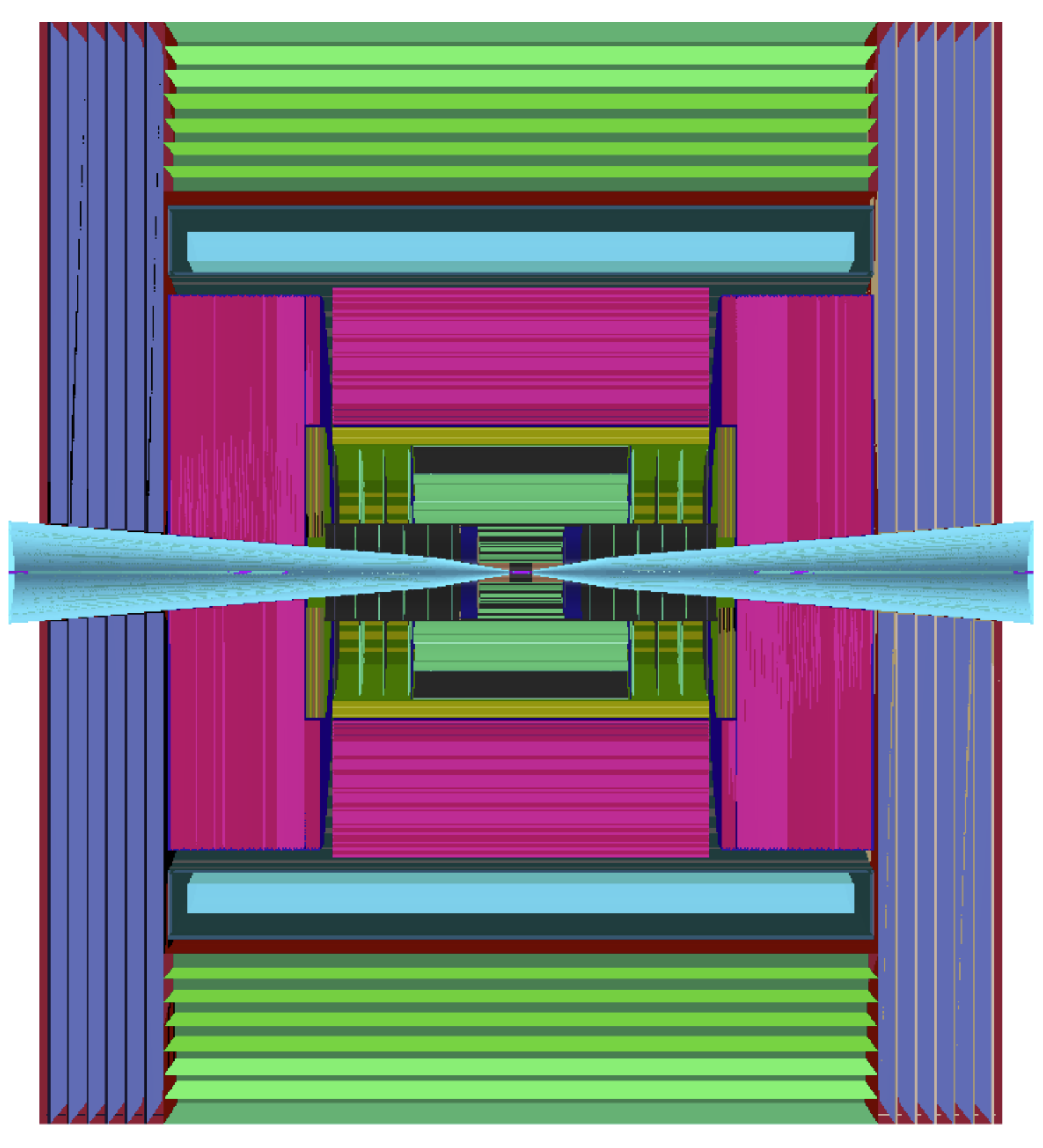}
\caption{Muon collider detector concept at $\sqrt{s} = 3$ TeV. From the innermost to the outermost regions, it includes the tracking system (green), the electromagnetic calorimeter (yellow), the hadronic calorimeter (magenta), the superconducting solenoid (light blue), the barrel (light green) and endcap (blue) muon detectors. The nozzles are shown in cyan.
\label{fig:geometry}}
\end{figure}

The detector, illustrated in \textbf{Figure~\ref{fig:geometry}}, consists of a tracking system and a calorimeter system, both immersed in a uniform solenoidal magnetic field, with an external muon detector.
From the innermost to the outermost regions with respect to the IP, it features the following components:
\begin{itemize}
\item[-] \textbf{Vertex detector} made of double layers of $25 \times 25 ~ \mu\mathrm{m}^2$ silicon pixels. It features four central barrel cylinders and four endcap disks on both sides of the barrel.
\item[-] \textbf{Inner tracker} with three barrel layers and seven endcap disks on each side made of silicon macropixels with a size of $50 ~ \mu\mathrm{m} \times 1 ~ \mathrm{mm}$.
\item[-] \textbf{Outer tracker} composed of three barrel layers and four endcap disks per side of silicon microstrips with a size of  $50 ~ \mu\mathrm{m} \times 10 ~ \mathrm{mm}$.
\item[-] \textbf{Electromagnetic calorimeter} (ECAL), composed of 40 alternating layers of tungsten absorber and silicon sensors as active material, for a total of 22 radiation lengths ($X_0$). The cell granularity is $5 \times 5$ mm$^2$.
\item[-] \textbf{Hadronic calorimeter} (HCAL), that has 60 alternating layers of steel absorber and scintillating-pads active material, for a total of 7.5 interactions lengths ($\lambda_I$) and with cells of size $30 \times 30$ mm$^2$.
\item[-] \textbf{Superconductive solenoid} generates a magnetic field of $B=3.57$ T; the magnetic field value has been chosen to be consistent with that used in the generation of the BIB sample.
\item[-] \textbf{Return yoke} equipped with resistive plate chambers (RPCs) for muon detection, with seven layers in the barrel and six layers in the endcap. Each cell has  an area of $30 \times 30$ mm$^2$.
\end{itemize}
The radius of the tracker barrel is 1.5 m, while the calorimeters barrel have outer radii of 1.7 m and 3.3 m, for ECAL and HCAL respectively. The total radial dimension of the detector, including the return yoke, is 6.5 m, and the detector length along the beam direction is $\pm 5.6$ m with respect to the IP. 

In order to study the impact of BIB on the physics performance, the interactions of these particles with the detector are simulated using the muon collider software framework~\cite{muc_framework}.
This software is a branched version of iLCSoft~\cite{ilcsoft}, a framework developed for electron-positron colliders.
The detector description is implemented with DD4hep~\cite{dd4hep} and the interaction of the BIB with the detector is simulated with \textsc{Geant4}~\cite{geant}. 
The BIB sample used to study the detector response was generated with MARS15 at $\sqrt{s}=1.5$ TeV. As discussed in \textbf{Section~\ref{sec:bck}}, beam-induced background samples generated with FLUKA at $\sqrt{s}=1.5$ and 3 TeV have become available only recently and show good compatibility.
The particles produced by the primary muon-muon collisions (\textit{i.e.} the physics signal) are generated with different programs like MadGraph5\_aMC@NLO~\cite{madgraph} or WHIZARD~\cite{whizard}. The interactions of final state particles with the detector are then simulated with \textsc{Geant4}.
The digitization of the detector hits and the reconstruction of the physics object are performed by the Marlin software~\cite{Marlin}.
Marlin also overlays the hits produced by the BIB particles onto the physics signal hits on an event-by-event basis.

By exploiting the detailed detector simulation, several strategies have been developed to mitigate the effects of BIB. Dedicated reconstruction algorithms have been employed to achieve high performance physics measurements. In the following sections, these techniques are described for the tracking and calorimeter systems (\textbf{Sections \ref{sec:tracking}} and \textbf{\ref{sec:calo}}, respectively). The discussion also covers the physics objects reconstruction performance achieved by combining information from the different sub-detectors (\textbf{Section \ref{sec:objects}})
\subsection{Beam-induced background mitigation in the tracking system}
\label{sec:tracking}
The BIB is responsible for a high hit occupancy in the tracking system, especially in the first layers of the vertex detector.
Hit densities of 3.68 mm$^{-2}$ and 0.51 mm$^{-2}$ are expected in the first and second pixel layers, respectively~\cite{muc_epjc}. These values are almost one order of magnitude higher than those predicted for the HL-LHC experiments. 
Nevertheless, it has to be remarked that the bunch crossing rate at the muon collider is expected to be much lower than that at LHC allowing more time for data processing.

The primary strategies for mitigating the BIB in the tracking system include: 
\begin{itemize}
\item[-]  Exploiting the hit times in the silicon sensors.
\item[-]  Utilizing directional information of the incoming particles.
\item[-]  Employing pulse shape analysis of the signal.
\item[-]  Analyzing the shape of hit clusters.
\end{itemize}
As explained in \textbf{Section \ref{sec:mdi}}, a significant fraction of the BIB is out-of-time with respect to the bunch crossing. 
Therefore, this component can be significantly reduced by requiring the hit times to be compatible with the timing of a particle coming from the IP.  
A time resolution of 30 ps is assumed for the vertex detector sensors, while a time resolution of 60 ps is considered for the inner and outer trackers. 

\begin{figure}[t]
\centering
\begin{minipage}{0.495\textwidth}
       \hspace{0.6cm}\textbf{\textsf{a}}\vspace{-0.15cm}

        \centering
        \includegraphics[width=1.0\textwidth]{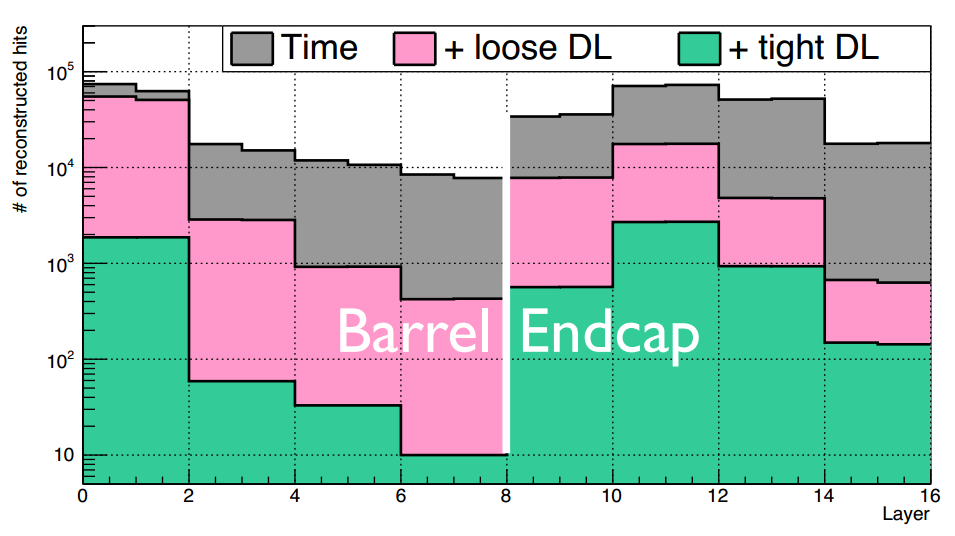}
    \end{minipage} \hfill
    \begin{minipage}{0.45\textwidth}
       \hspace{0.6cm}\textbf{\textsf{b}}\vspace{-0.4cm}
       
        \centering
        \includegraphics[width=1.0\textwidth]{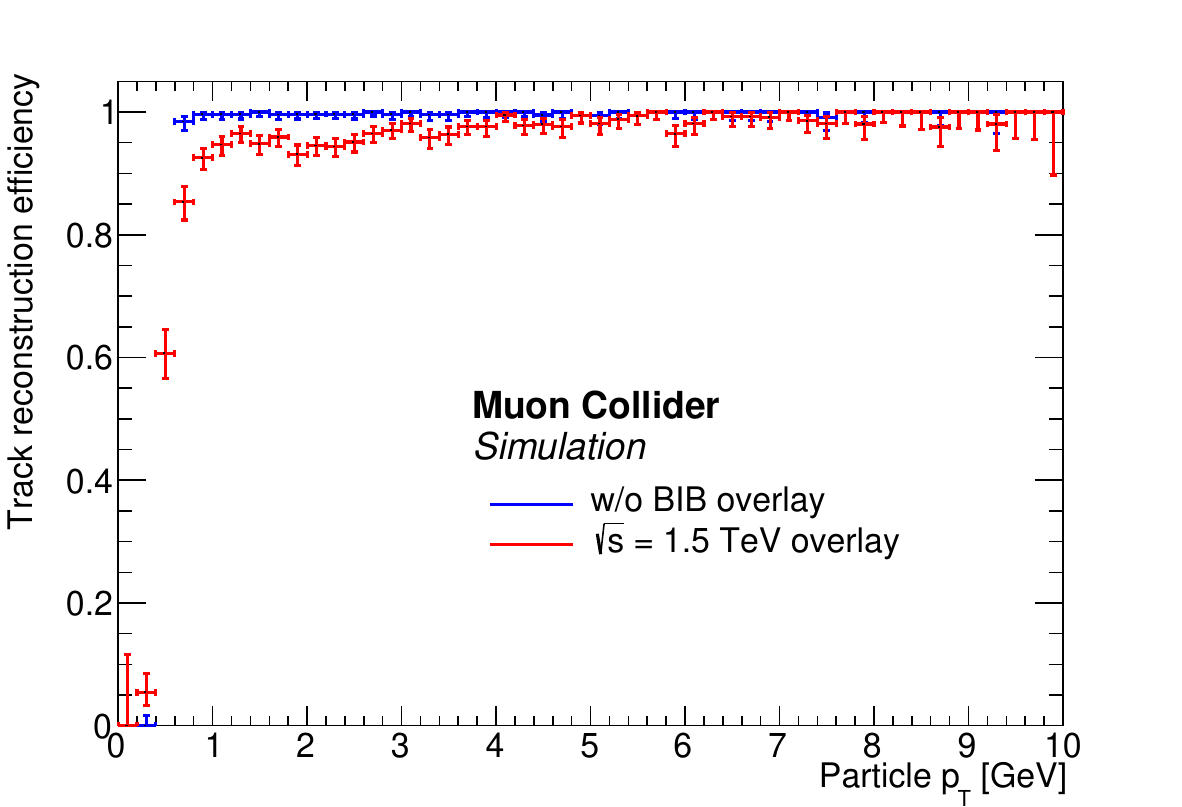}
    \end{minipage}
\caption{\textit{(a)} Hit density as a function of the vertex detector layer, with the reduction obtained with time requirements and double layer (DL) filters. \textit{(b)} Muon track reconstruction efficiencies as a function of the $p_{\mathrm{T}}$, with and without the BIB overlaid \cite{muc_epjc}. The double layer filters are not applied in this plot.
\label{fig:tracking}}
\end{figure}

Most of the BIB particles arriving at the detector originate far from the IP, so their incoming direction does not point directly to the interaction region. 
The double-layer configuration of the vertex detector can be exploited to estimate the direction of the particle by correlating the hit positions on adjacent layers. This strategy is similar to what is being developed for the upgrade of the CMS pixel detector, mainly for aiding the trigger system \cite{cmspixel}. Again, it is important to notice how the muon collider detector R\&D could benefit from synergies with HL-LHC detectors. 
The effective hit occupancy reduction in the vertex detector, achieved by applying requirements on time and direction, is presented in \textbf{Figure \ref{fig:tracking}\textit{a}}. 
Notably, the hit density is reduced by more than an order of magnitude when using the double layer requirement. A further reduction can be achieved by considering the pulse shape at the sensor level and the shape of the hit clusters, but these studies are ongoing.

The hits that survive the BIB suppression are then utilized as input for the track reconstruction procedure.
The primary algorithm employed for this purpose makes use of the Combinatorial Kalman Filter, often referred to as the CKF algorithm~\cite{CKF}. This algorithm reconstructs the trajectory of charged particles in the magnetic field (tracks), allowing the measurement of their transverse momentum ($p_{\mathrm{T}}$).
The tracking algorithm was tested on samples of muons generated at the primary vertex. 
\textbf{Figure \ref{fig:tracking}\textit{b}} shows the track reconstruction efficiency as a function of muon $p_{\mathrm{T}}$. To understand the impact of the BIB, these efficiencies are compared with those obtained without the BIB overlay. It is evident that the BIB leads to a drop in efficiency at low $p_{\mathrm{T}}$, below 4 GeV, while the tracking efficiency remains close to 1 for muons with higher $p_{\mathrm{T}}$.
The estimated uncertainty on the transverse momentum is $\Delta p_{\mathrm{T}} / p^2_T \approx 1 \cdot 10^{-1}\ (5 \cdot 10^{-3})$ GeV$^{-1}$ for muons with momentum $p=1\ (100)$ GeV and $\theta = 13^{\circ}$, and $\Delta p_{\mathrm{T}} / p^2_T \approx 5 \cdot 10^{-2}\ (4 \cdot 10^{-5})$ GeV$^{-1}$ for muons with $p=1\ (100)$ GeV and $\theta = 89^{\circ}$ \cite{muc_epjc}.
\subsection{Beam-induced background mitigation in the calorimeter system}
\label{sec:calo}
\begin{figure}[t]
\begin{minipage}{0.5\textwidth}
       \hspace{0.6cm}\textbf{\textsf{a}}\vspace{-0.0cm}

        \centering
        \includegraphics[width=1.0\textwidth]{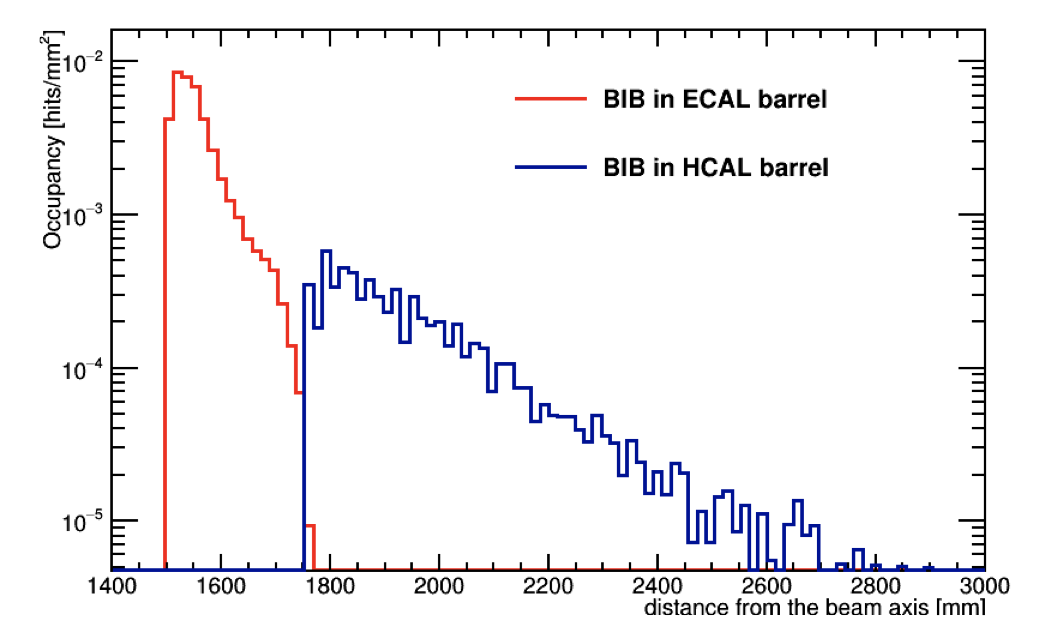}
    \end{minipage} 
    \begin{minipage}{0.46\textwidth}
       \hspace{0.6cm}\textbf{\textsf{b}}\vspace{-0.4cm}
       
        \centering
        \includegraphics[width=1.0\textwidth]{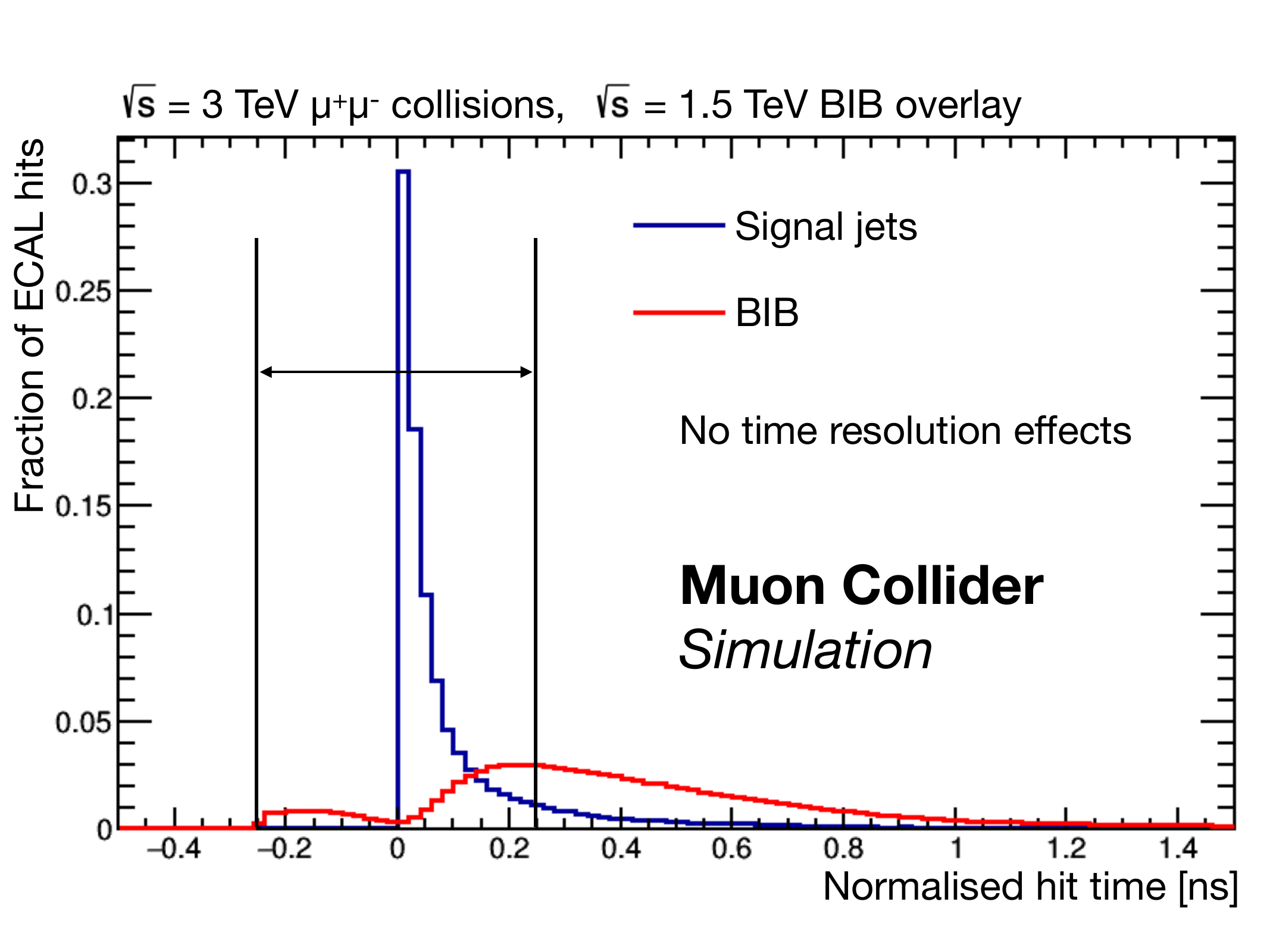}
    \end{minipage}
    \centering
    \begin{minipage}{0.6\textwidth}
       \hspace{0.8cm}\textbf{\textsf{c}}\vspace{-0.3cm}
       
        \centering
        \includegraphics[width=1.0\textwidth]{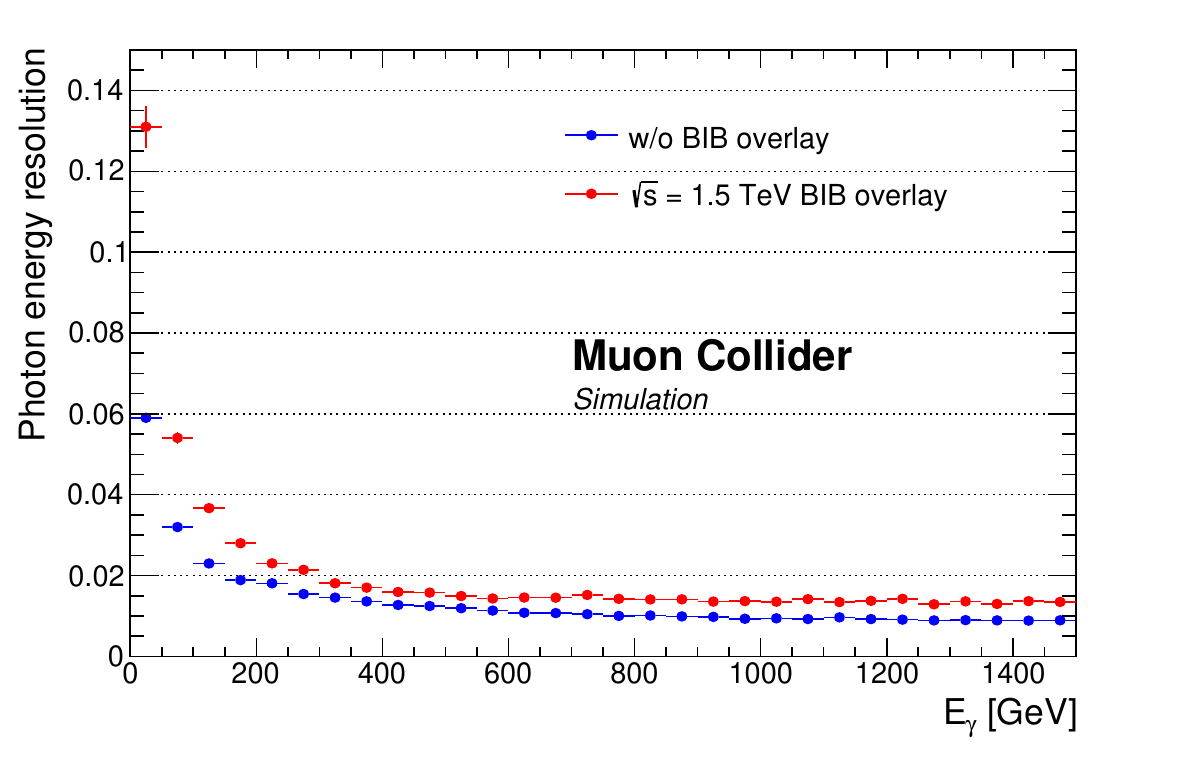}
    \end{minipage} \hfill
\caption{\textit{(a)} Calorimeter occupancy as a function of the radial distance from the beam axis. \textit{(b)} Distributions of the ECAL hit times with respect to the bunch crossing for signal (hadronic jets) and BIB particles, both normalized to unity area. \textit{(c)} Photon energy resolution for low- and high-energy photons, obtained with and without the BIB overlaid \cite{muc_epjc}.
\label{fig:calorimeter}}
\end{figure}
At the muon collider, the internal surface of the electromagnetic calorimeter (ECAL) is exposed to a flux of about 300 BIB particles per cm$^2$ at each bunch crossing. The majority of these particles are photons ($96\%$) with an average energy of 1.7 MeV and the remaining $4\%$ are neutrons. 
The calorimeter occupancy as a function of the radial distance from the beam axis is shown in \textbf{Figure \ref{fig:calorimeter}\textit{a}}.
The occupancy in the first layers of the ECAL barrel is in the order of $10^{-2}$ hits/mm$^2$ (corresponding on average to 1 hit per cell), but it quickly decreases down to 10$^{-4}$ hits/mm$^2$ in the last ECAL layers.
Neutrons that don't interact with the ECAL reach the HCAL where, on the first layer, the occupancy is approximately one order of magnitude lower than that of the first ECAL layer~\cite{muc_epjc}. 
The following features can be exploited to suppress the BIB in the calorimeter:
\begin{itemize}
\item[-] Arrival time of the particles in the calorimeter cells.
\item[-] Calorimeter granularity.
\item[-] Energy thresholds to reject the soft component.
\item[-] Calorimeter segmentation for the measurement of the longitudinal energy distribution.
\end{itemize}
Similarly to the tracking system, time information plays an important role also in the calorimeters. As shown in \textbf{Figure \ref{fig:calorimeter}\textit{b}},
the times of a consistent fraction of the BIB hits of the ECAL are outside the time window of the physics signal (\textit{e.g.} hadronic jets). 
Assuming a calorimeter hit time resolution of approximately  100 ps, it is possible to apply a time window requirement of $\pm$250 ps, which efficiently suppresses a significant portion of BIB hits.

A high granularity and a fine longitudinal segmentation of the calorimeter could help in separating the showers initiated by signal particles from those due to the BIB. This is the reason why cells of $5 \times 5$ mm$^2$ are considered for ECAL. 

To suppress the low energy deposits associated with the BIB, a threshold of 2 MeV is set on the energy of the calorimeter cells. This threshold, while relatively high, represents a compromise between performance, BIB suppression and computation time. Nonetheless, new handles can be explored in the future to lower this threshold and enhance performance. For instance, the analysis of the longitudinal distribution of the deposited energy in the calorimeter can be leveraged to differentiate signal from BIB showers, as distinct profiles are expected.

The photon energy resolution achieved with the current calorimeter configuration and the aforementioned requirements is presented in \textbf{Figure \ref{fig:calorimeter}\textit{c}}. The resolution is approximately $13\%$ for low energy photons below 50 GeV, and around $1.5\%$ for photons with energy above 400 GeV. By comparing the resolution with that obtained without the BIB overlaid, it becomes evident that the BIB leads to a degradation in performance.

\subsection{Physics objects reconstruction with particle flow}
\label{sec:objects}
Several physics objects are crucial for the success of the muon collider physics program. 
The reconstruction and identification of muons, photons, electrons and jets have been studied with the detailed detector simulation. As explained in the following sections, these particles are present in the final states of the most important Higgs decay channels and may be part of the signatures for new physics searches.

The reconstruction of these physical objects involves combining information from all the detector sub-systems.  Tracks, calorimeter clusters, and muon detector hits are combined to achieve  optimal performance in terms of identification efficiency, background rejection, and momentum resolution. For this purpose, a particle flow algorithm, PandoraPFA~\cite{pandora}, is employed.
PandoraPFA is an algorithm based on modern particle flow techniques and has been used for object reconstruction tasks. Initial studies have demonstrated its effectiveness in the muon collider environment, although its initial configuration was  optimized for electron-positron colliders. It is evident that there is room for further improvements and, in the future, a dedicated optimization of the PandoraPFA settings for the muon collider can be pursued.

\begin{figure}[t]
\centering
\begin{minipage}{0.45\textwidth}
       \hspace{0.7cm}\textbf{\textsf{a}}\vspace{-0.3cm}

        \centering
        \includegraphics[width=1.0\textwidth]{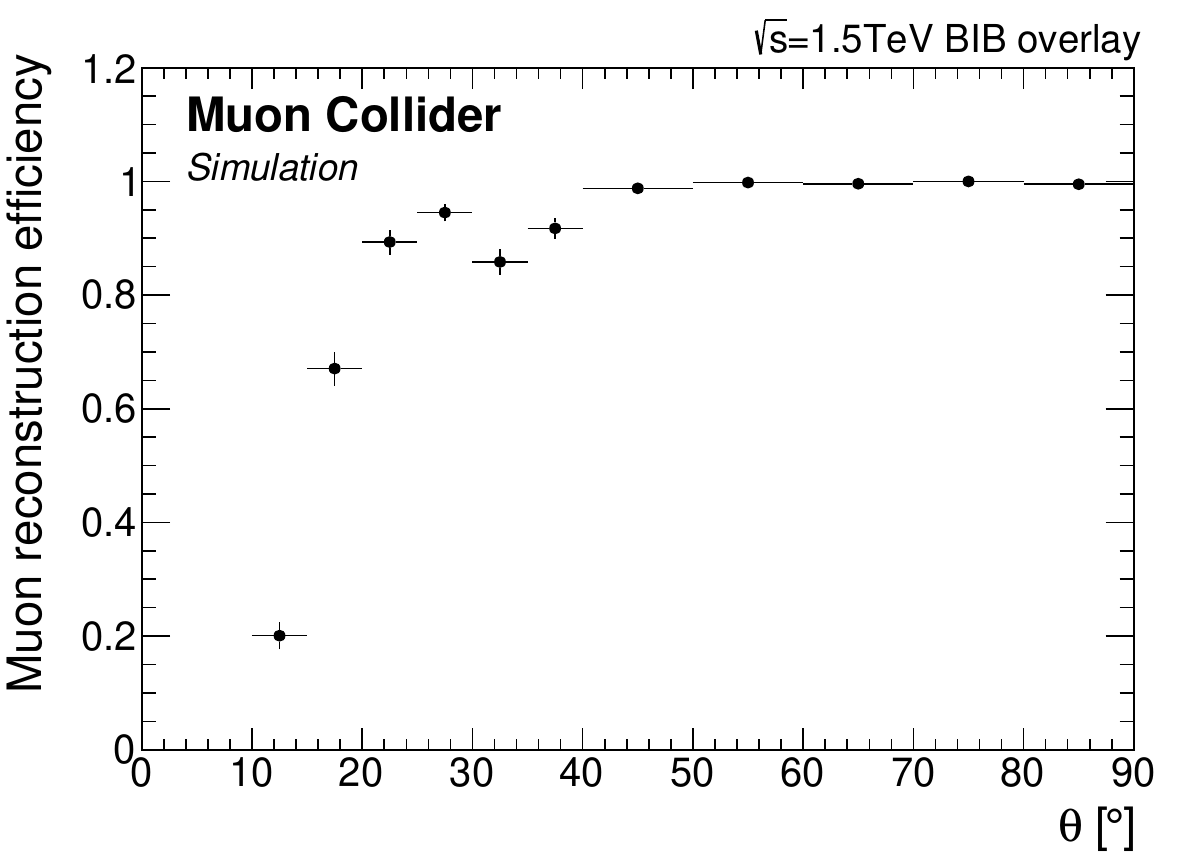}
    \end{minipage} \hfill
    \begin{minipage}{0.51\textwidth}
       \hspace{0.7cm}\textbf{\textsf{b}}\vspace{-0.25cm}
       
        \centering
        \includegraphics[width=1.0\textwidth]{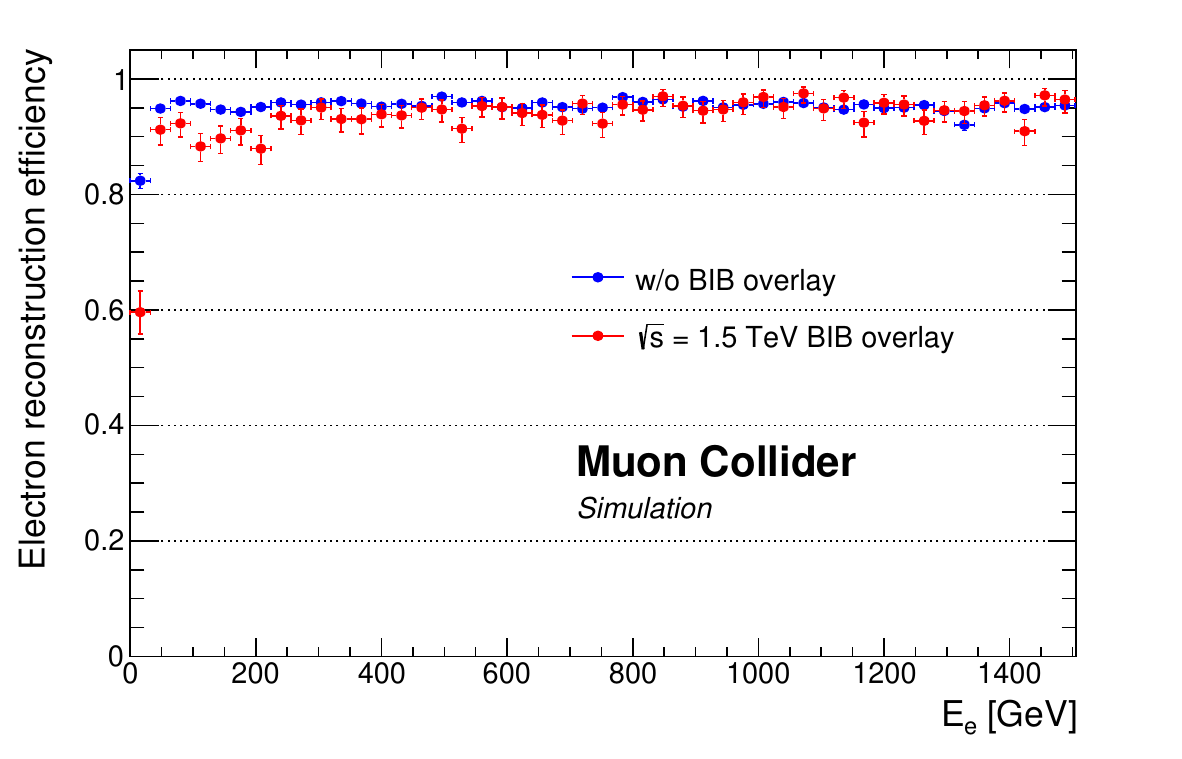}
    \end{minipage}
\begin{minipage}{0.495\textwidth}
       \hspace{0.9cm}\textbf{\textsf{c}}\vspace{-0.2cm}

        \centering
        \includegraphics[width=1.0\textwidth]{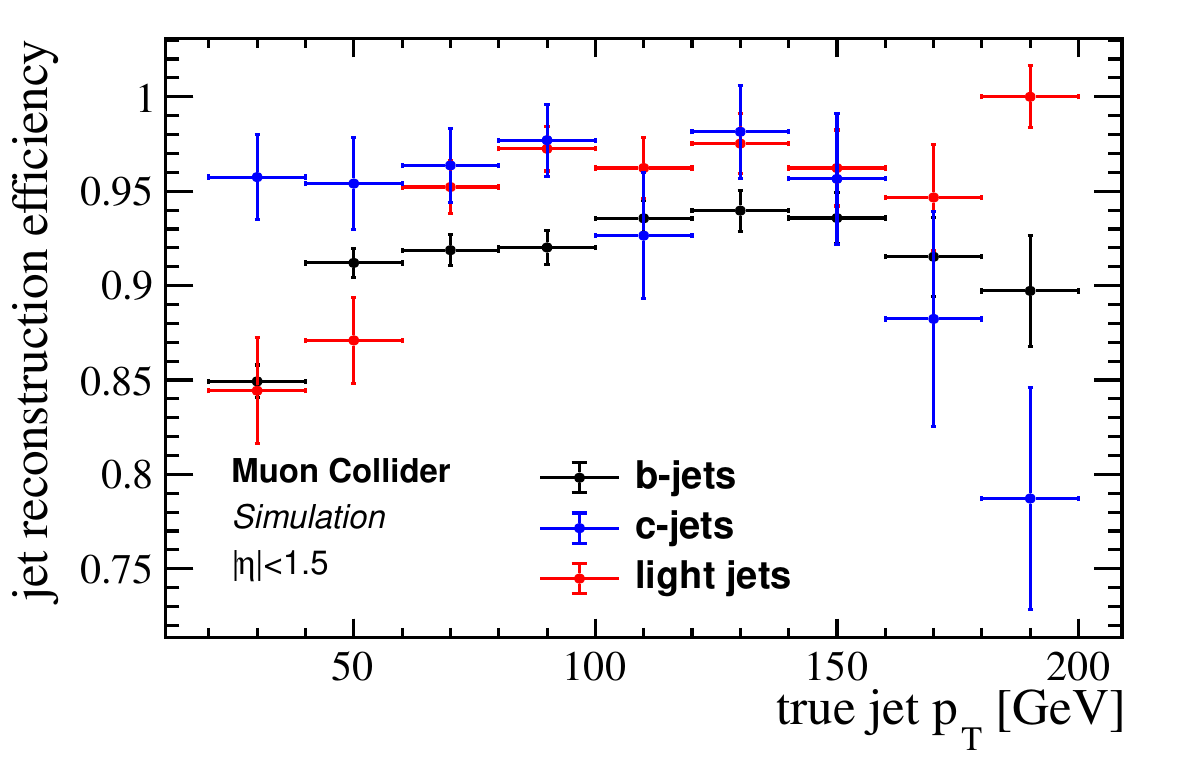}
    \end{minipage} \hfill
    \begin{minipage}{0.48\textwidth}
       \hspace{0.6cm}\textbf{\textsf{d}}\vspace{-0.25cm}
       
        \centering
        \includegraphics[width=1.0\textwidth]{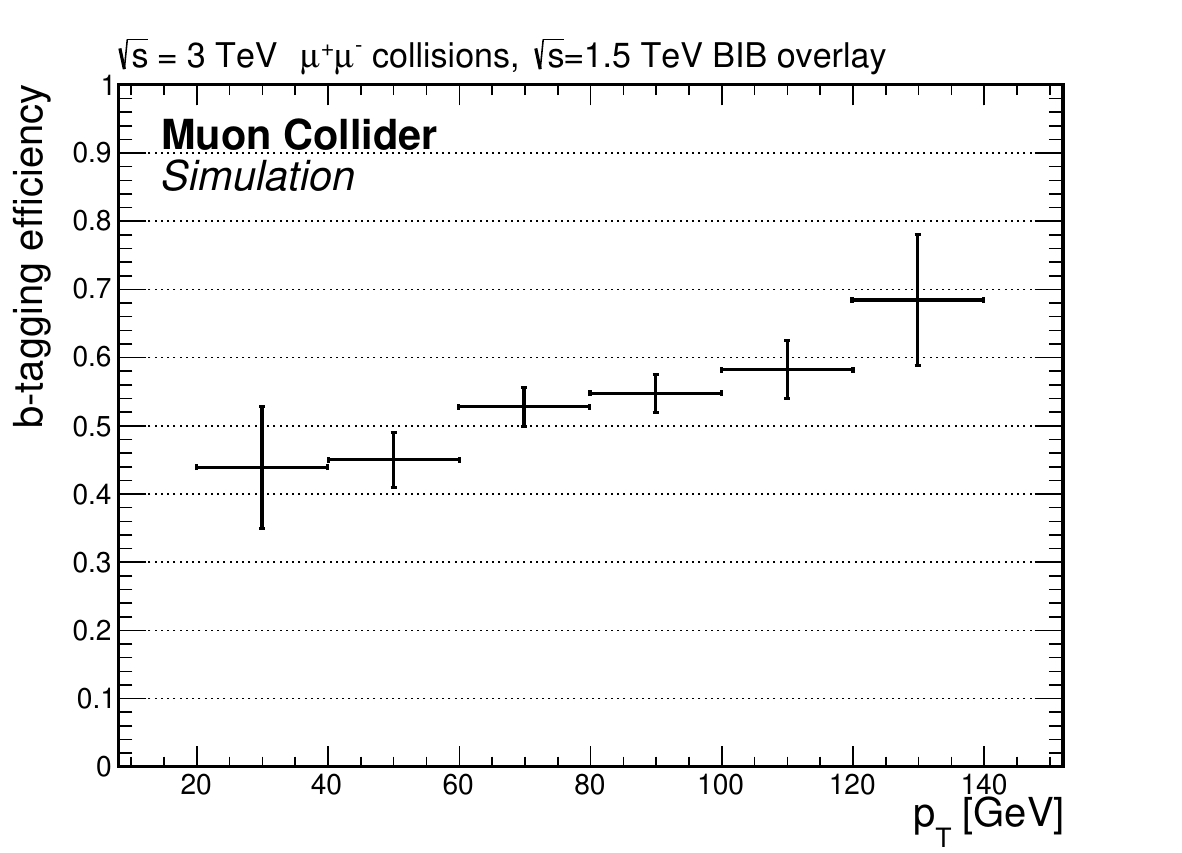}
    \end{minipage}
\caption{\textit{(a)} Muon reconstruction efficiency as a function muon $p_{\mathrm{T}}$. \textit{(b)} Electron reconstruction efficiency as a function of the electron energy, with and without the BIB overlay. \textit{(c)} Jet reconstruction efficiency as a function of jet $p_{\mathrm{T}}$, for different jet flavours. \textit{(d)} $b$-tagging probability as a function of the jet $p_{\mathrm{T}}$. The above figures have been reproduced from~\cite{muc_epjc}.
\label{fig:particleflow}}
\end{figure}

The reconstruction of the following physics objects has been studied, along with the resulting performance:
\begin{itemize}
\item[-] \textbf{Muons} are identified by matching tracks with hits in the muon detector system. This sub-detector is not significantly affected by the BIB,
except for the forward region with respect to the beam direction. In \textbf{Figure \ref{fig:particleflow}\textit{a}}, the muon reconstruction and identification efficiency is presented as a function of the muon production angle with respect to the beamline. It is evident that the low angle region suffers from the BIB presence, while the central region exhibits high efficiency.

\item[-] \textbf{Electrons} and \textbf{photons} are reconstructed by matching tracks with ECAL clusters. Isolated clusters are classified as photons, while clusters matched with a track are classified as electrons.
The energy of electrons and photons is corrected to take into account inefficiencies and radiation losses. The electron reconstruction efficiency as a function of the energy is presented in \textbf{Figure \ref{fig:particleflow}\textit{b}}: it reaches around 95$\%$ for high energy, but the typical efficiency drop at low energy caused by the BIB is evident.

\item[-] \textbf{Hadronic jets}: particles reconstructed by PandoraPFA are given in input to the jet clustering algorithm, that groups together particles belonging to the same fragmentation process, by exploiting their correlations. The jet algorithm used is the $k_t$ algorithm \cite{jetkt1,jetkt2} with a radius parameter of 0.5. The resulting jet energy is corrected to recover the losses due to particles escaping the detector, detector inefficiencies, and also for the BIB contamination. The jet reconstruction algorithm has been tested with simulated samples of different jet flavours: gluons and $b$, $c$, $u$, $d$, and $s$ quarks. The reconstruction efficiencies are illustrated in \textbf{Figure \ref{fig:particleflow}\textit{c}}. Despite the presence of BIB throughout the detector, the efficiency ranges between 80$\%$ and 95$\%$, with a negligible fake jet probability.

\item[-] \textbf{Heavy flavour jets}: in order to identify the jets originating from heavy quarks, displaced secondary vertices compatible with the decays of $b$- and $c$-hadrons are reconstructed by combining tracks. A jet is identified as a heavy-flavour jet if one of these secondary vertices is reconstructed within the jet cone. The $b$-jet identification probability, called $b$-tagging efficiency, is shown in \textbf{Figure \ref{fig:particleflow}\textit{d}}. It is of approximately 45$\%$ at low $p_{\mathrm{T}}$ (20 GeV) and increases to 70$\%$ at 120 GeV. This level of performance is comparable to what is currently achieved at hadron colliders.
At lepton colliders, a higher efficiency is expected. The current algorithm is mainly limited by the presence of the BIB. In the future, artificial-intelligence-based methods  will be implemented to minimize the effects of BIB and further enhance efficiency.
\end{itemize}
%

\subsection{Luminosity measurement}
\label{sec:lumi}
Accurate knowledge of the absolute luminosity is crucial in collider experiments, as it is essential for determining the production cross sections. Uncertainty below $1\%$ has been  achieved by the ATLAS~\cite{atlas-lumi} and Belle II~\cite{belle-lumi} experiments. 
In general, the luminosity measurement methods rely on the direct proportionality between the rates of some chosen processes observed in the detector and luminosity.
Two different approaches are typically used for hadron and lepton colliders:

\begin{itemize}
    \item [-] At hadron colliders, the rate $R$ of a given process is measured and related to the luminosity by $\mathcal{L}=R/\sigma_{vis}$.
    The visible cross section $\sigma_{vis}$ is defined as $\sigma_{vis}=2\pi\,A_{eff}/(N_1 N_2)$, where $N_{1,2}$ is the average number of particles per bunch provided 
    by the machine and $A_{eff}$ is the effective transverse area of the luminous region.
    In this method, a precise calibration of the visible cross section is crucial. At the LHC, it is achieved with dedicated van der Meer 
    scans~\cite{luminometer}, in which the beams are displaced against each other.

    \medskip
    \item[-] Experiments at lepton colliders count the number of events $N_{ev}$ of a reference process, whose cross section $\sigma$ is 
    sufficiently high and theoretically know with very good precision, and determine the luminosity by $\mathcal{L}_{int} = N_{ev}/(\epsilon\, \sigma)$, where $\epsilon$ is the
    total selection efficiency for the considered events. 
    Electron-positron experiments exploit the abundant $e^+e^- \to e^+e^-$ elastic scattering (Bhabha scattering), for which the cross section is 
    known with a precision below a percent at $\sqrt{s}=1$-10 GeV~\cite{bhabha_cross_section}.
    
    \end{itemize}
In both cases, the rate of the reference processes is measured by dedicated detectors, called luminometers, that are installed at very small 
polar angles on both sides of the interaction point, where the observed rates are higher.
In a muon collider experiment, the presence of the nozzles hinders the installation of the luminometers and an alternative technique must be utilized
to determine the luminosity.
A possible approach could be to count muons scattered elastically at large angles, similar to what was done in KLOE~\cite{kloe-lumi}.

A preliminary study verified the feasibility of such a method~\cite{lumicarlo}.
The Pythia Monte Carlo generator~\cite{pythia} was used to produce a sample of $\mu^+ \mu^- \to \mu^+ \mu^-$ Bhabha events with an emission angle 
of $30^\circ <\theta_\mu < 150^\circ$ with respect to the beam direction at a center-of-mass energy of 1.5 TeV.
The sample was processed with the detector simulation and the muon collider reconstruction software.
The event requirements are quite simple: two opposite-charge muons with a transverse momentum $p_T >130$ GeV and 
an invariant mass $1440<m_{\mu\mu}<1560$ GeV. Background contributions from other physics processes are found to be negligible
as well as possible effects from the beam-induced background.
The statistical uncertainty on the integrated luminosity estimated at $\sqrt{s}=1.5$ TeV is:
\begin{equation}
  \frac{\Delta N_{Bhabha}}{N_{Bhabha}} = \frac{1}{\sqrt{N_{Bhabha}}} = 0.002\ .
\end{equation}
This approach requires a precise theoretical prediction for the radiatively corrected Standard Model $\mu$-Bhabha cross section at high center of mass energies, which should be straightforward to obtain.
Assuming this, the method seems promising.
In the future, other processes could be studied in order to explore which ones could offer competitive luminosity measurements. An example is the $\mu^+ \mu^- \rightarrow W^+ W^-$ production, which has a large cross sections at high collision energies.
%
\section{THE GUARANTEED PHYSICS DISCOVERY: FULL DETERMINATION OF THE HIGGS BOSON PROPERTIES} 
\label{sec:higgs}
The Higgs boson plays a fundamental role in the Standard Model and could hold the key to a deeper understanding of the universe. Its observation was crucial in establishing the mechanism of electroweak symmetry breaking and mass generation~\cite{PDG}.
Precisely determining the Higgs boson properties and understanding its nature are therefore of
the utmost importance in the particle physics field and are central to the physics programs of
all the proposed next-generation colliders.

\begin{figure}[t]
  \begin{minipage}{0.495\textwidth}
    \hspace{0.95cm}\textbf{\textsf{a}}\vspace{-0.2cm}

    \centering
    \includegraphics[width=0.93\textwidth]{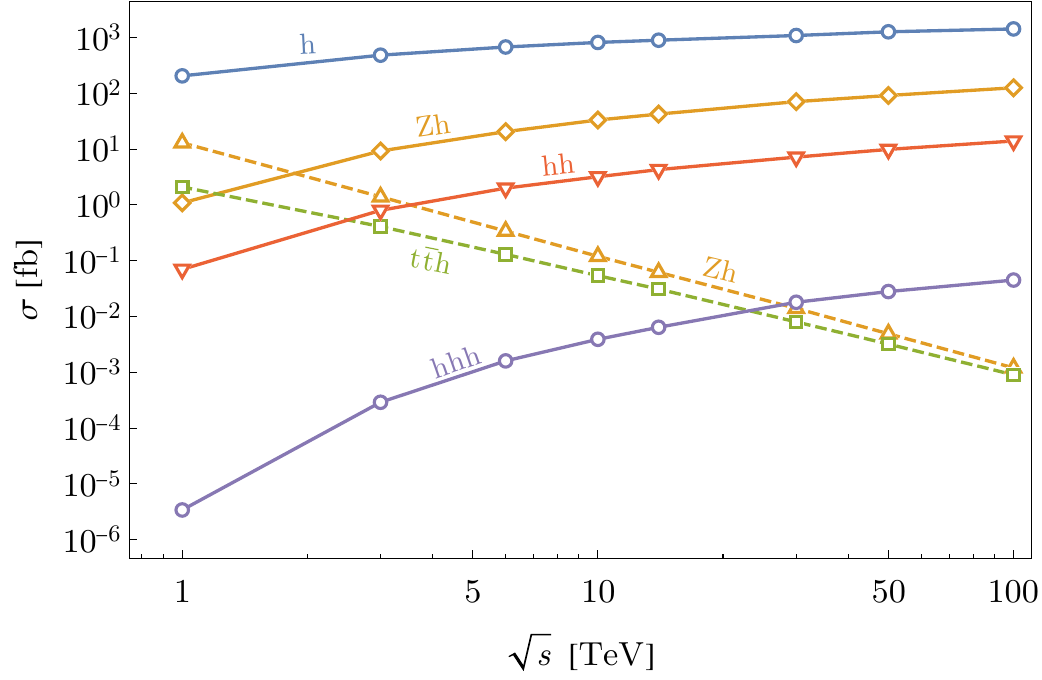}
  \end{minipage} \hfill
  \begin{minipage}{0.495\textwidth}
    \hspace{0.9cm}\textbf{\textsf{b}}\vspace{-0.4cm}

    \centering
    \includegraphics[width=\textwidth]{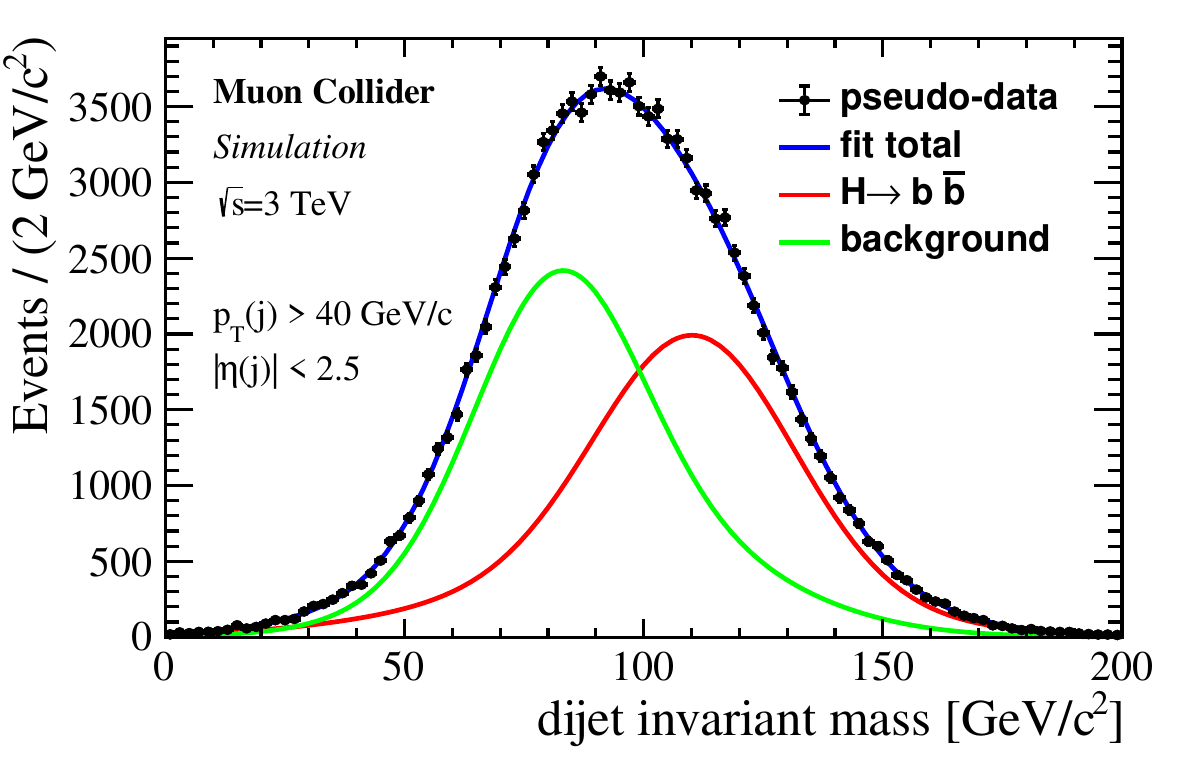}
  \end{minipage}
  \begin{minipage}{0.495\textwidth}
    \hspace{0.9cm}\textbf{\textsf{c}}\vspace{-0.4cm}

    \centering
    \includegraphics[width=\textwidth]{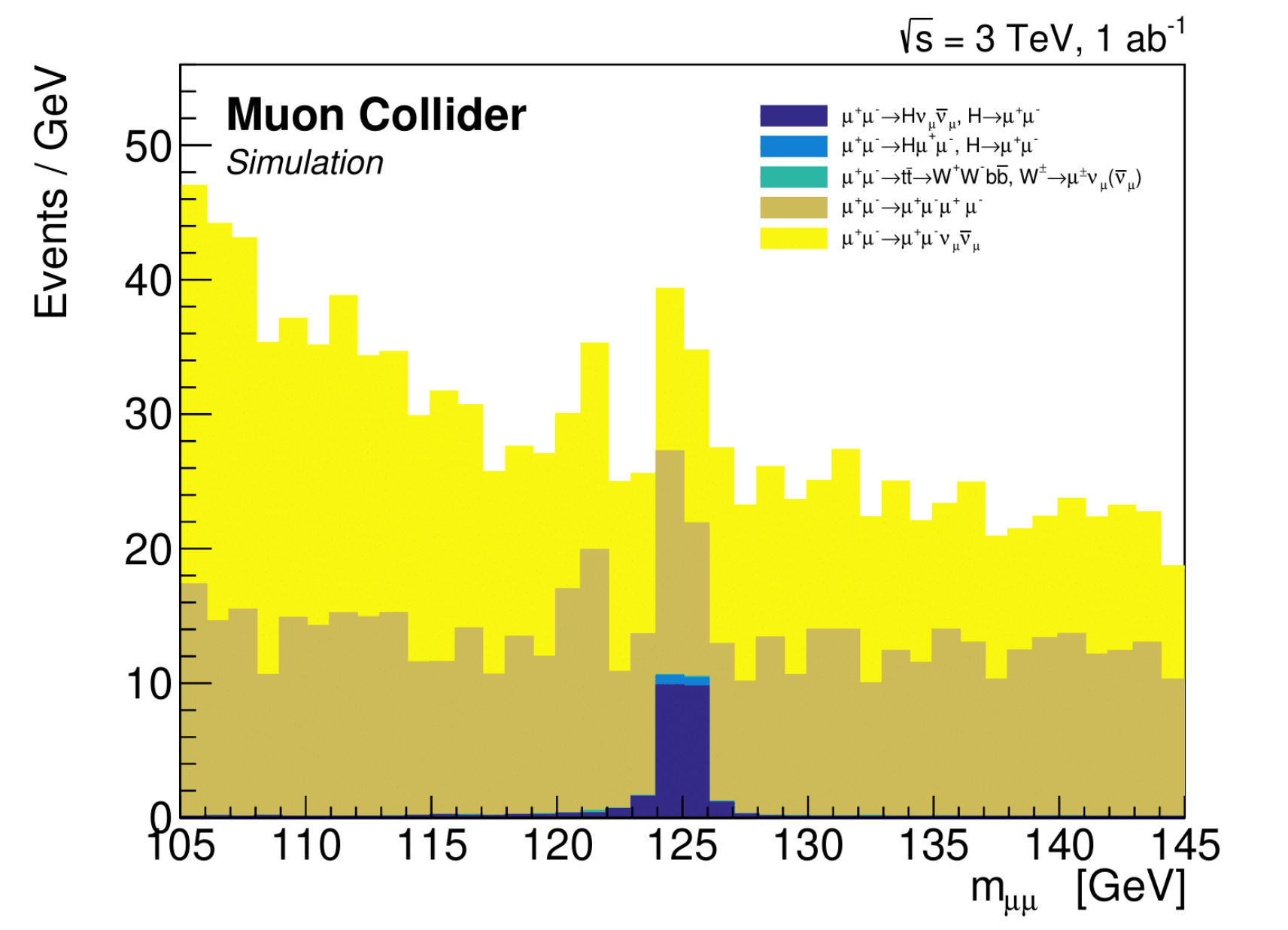}
  \end{minipage}\hfill
  \begin{minipage}{0.495\textwidth}
    \hspace{0.9cm}\textbf{\textsf{d}}\vspace{-0.4cm}

    \centering
    \includegraphics[width=\textwidth]{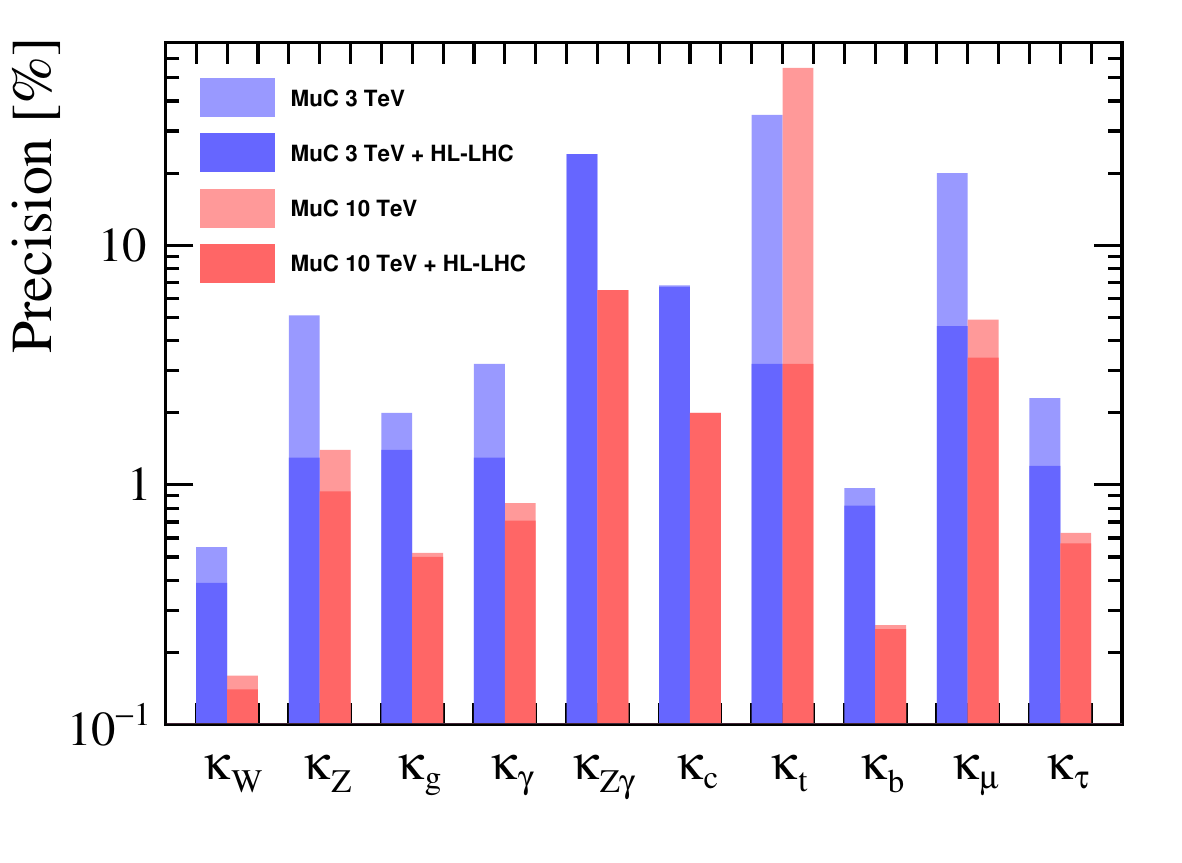}
  \end{minipage}
  \caption{({\it a}) Cross sections of the main Higgs boson production channels as a function of
    the muon collider center-of-mass energy $\sqrt{s}$, from Ref.~\cite{smasher_guide}. The cross 
    sections of processes mediated by vector-boson fusion increase with energy, while those mediated 
    by $s$-channel $\mu^+\mu^-$ annihilation decrease. 
    ({\it b}) Invariant mass of the two jets reconstructed in the $H\to b\bar{b}$ channel:
    the reconstructed mass peak of the Higgs boson is indicated in red, while the background, dominated 
    by $Z^0\to b\bar{b}$ and $c\bar{c}$, is in green.
    ({\it c}) Invariant mass of the two muons in the $H\to\mu^+\mu^-$ analysis: the backgrounds
    (in yellow and brown) are stacked on top of the signal (dark blue).
    ({\it d}) Precision on the Higgs boson couplings at a 3 TeV and 10 TeV muon collider, with 
    a dataset of 1 ab$^{-1}$ and 10 ab$^{-1}$, respectively, in the assumption that there are no 
    BSM decays of the Higgs boson. The light-colored histograms indicate the expected precision of muon collider alone, while the dark-colored ones show the precision
    when combining with the HL-LHC. The plotted values are taken from~\cite{patrick1}.
    \label{fig:Higgs}}
\end{figure}

A high energy muon collider represents an ideal machine to study in detail the Higgs boson properties,
since it is expected to provide large samples of single Higgs bosons as well as unprecedented samples
of multi-Higgs bosons in a relatively clean experimental environment.
The cross sections of the most important production channels are shown in \textbf{Figure~\ref{fig:Higgs}\textit{a}}
as a function of the muon collider center-of-mass energy $\sqrt{s}$.
At multi-TeV lepton collisions, Higgs bosons are produced via two fundamental processes:
the vector boson fusion and the muon-antimuon annihilation.
The cross sections of the latter decrease as $1/s$ with increasing collision energies,
whereas the former displays cross sections that increase logarithmically with $s$ and
eventually become dominant at high energies.

A 3 TeV muon collider with a dataset of 1 ab$^{-1}$ is expected to produce a sample of
approximately $5.5\times 10^{5}$ single Higgs bosons and a sample of about 1000 double Higgs
events.
A muon collider operating at 3 TeV would undoubtedly advance our understanding of the Higgs 
boson's properties significantly, but it is at higher energies that the full potential of a 
muon collider emerges.
At a 10 TeV collider with a dataset of 10 ab$^{-1}$, approximately 9.3 million
single Higgs bosons and a sample of about 38000 double-Higgs events are expected. 
The 1 ab$^{-1}$ and 10 ab$^{-1}$ datasets correspond to the integrated luminosity targets set in the IMCC's studies~\cite{muc_epjc}.

%
%
\subsection{The Higgs boson couplings determination}
A first campaign of studies to estimate the precision on the Higgs boson cross sections in different final states  
was carried out at $\sqrt{s} = 3$ TeV with a detailed detector simulation
that includes the beam-induced background. Given the unique nature of such a background, it was not
possible to model or parameterize its impact on the detector performance, and hence the physics reach,
relying on prior experience with machine backgrounds at current or past colliders.

The physics objects described in \textbf{Section~\ref{sec:objects}} were used
to reconstruct the Higgs boson decays into the final states $f = b\bar{b}$, $WW^\ast$,
$ZZ^\ast$, $\gamma\gamma$, and $\mu^+\mu^-$ to estimate the statistical sensitivity on the
production cross sections multiplied by the decay branching ratios: $\sigma_H \times BR(H\to f)$.

The Higgs and physics background samples were generated at leading order with WHIZARD or MadGraph5\_aMC@NLO
and Pythia8 was used for the final state hadronization. The samples were then processed with the detailed detector
simulation and reconstructed with the muon collider software.
The results are outlined below. A dataset of 1 ab$^{-1}$ and a single interaction point are assumed.
A natural reference to compare to are the corresponding results by CLIC at $\sqrt{s} = 3$ TeV~\cite{clic}
with the caveat that CLIC is assuming a dataset of 2 ab$^{-1}$ and the number of reconstructed final
states for each Higgs boson decay mode are in some cases different.
\begin{description}
\item[$\bm{H\to b\bar{b}:}$]
  The $b\bar{b}$ channel~\cite{buonincontri} is reconstructed selecting two central high-$p_T$ jets satisfying: $p_T^{\textsuperscript{jet}} > 40$ GeV
  and $|\eta_{\textsuperscript{jet}}| < 2.5$.
  Once the $b$-flavor identification is applied to both jets, the only significant background that survives is
  from $Z^0$ boson decays into $b\bar{b}$ and $c\bar{c}$ pairs. A total number of 59500 signal events and 65400
  background events are estimated.  
  \textbf{Figure~\ref{fig:Higgs}\textit{b}} shows the reconstructed dijet invariant mass. 
  The dijet mass resolution is critical to separate the $H$ and $Z^0$ peaks. The current value of the Higgs boson mass resolution is of about $18\%$, being dominated by BIB effects, \textit{i.e.} the energy thresholds 
  set on the calorimeter hits (see \textbf{Section~\ref{sec:calo}}).
  The estimated sensitivity on $\sigma_H \times BR(H\to b\bar{b})$ is 0.75\%, to be compared with 0.3\% of CLIC.
  
  \medskip
  
\item[$\bm{H\to WW^\ast:}$]
  The $WW^\ast$ channel~\cite{castelli} is reconstructed in the semileptonic final state $qq^\prime \mu\nu_\mu$,
  which provides a good signal-to-background ratio.
  Events are selected with at least two reconstructed jets, having $p_T^{\textsuperscript{jet}} > 20$ GeV and
  $|\eta_{\textsuperscript{jet}}| < 2.5$, and one muon with $p_T^\mu > 10$ GeV and $10^\circ < \theta_\mu < 170^\circ$.
  The estimated number of Higgs candidates is 2430 over a background of 2600 events.
  The resulting statistical uncertainty on the production cross section is 2.9\%.
  Reconstructing the full hadronic channel and the semileptonic channels with both muons and electrons, CLIC achieves a sensitivity of 0.7\%.

  \medskip
  
\item[$\bm{H\to ZZ^\ast:}$]
  The $ZZ^\ast$ channel~\cite{sestini} is reconstructed in the semileptonic final state $q\bar{q} \mu^+\mu^-$.
  Events are selected with at least two reconstructed jets and two opposite-charge muons, having
  $p_T^{\textsuperscript{jet}} > 15$ GeV and $30^\circ < \theta_\textsuperscript{jet} < 150^\circ$,
  $p_T^\mu > 10$ GeV and $10^\circ < \theta_\mu < 170^\circ$.  
  The sensitivity estimated on the cross section with 55 signal events and 39
  background events is 17\%.
  CLIC analyses also the semileptonic channel with electrons and gets 3.9\%.

  \medskip
  
\item[$\bm{H\to \gamma\gamma:}$]
  The $\gamma\gamma$ analysis~\cite{casarsa} searches for events with at least two reconstructed photons,
  featuring $E_\gamma > 15$ GeV, $p_T^\gamma > 10$ GeV and $10^\circ < \theta_{\gamma} < 170^\circ$.
  The reconstruction of high energy photons is not significantly affected by the BIB. In fact, the diphoton
  invariant mass exhibits a very good resolution of 3.2 GeV.
  The estimated numbers of signal and background events are 396 and 484, respectively, which result in a relative uncertainty on the production cross section of 7.6\%, to be compared with CLIC's 10\%.

  \medskip  
  
\item[$\bm{H\to \mu^+\mu^-:}$]
  The $\mu\mu$ channel~\cite{montella} is reconstructed selecting events with two opposite-charge muons in the kinematical region
  $p_T^\mu > 5$ GeV and $10^\circ < \theta_\mu < 170^\circ$.
  The BIB impact on the reconstruction of muons is found to be negligible. The reconstructed dimuon mass is shown in \textbf{Figure~\ref{fig:Higgs}\textit{c}}, where the signal peak has a width
  of 0.4 GeV. After the final selection, 26 signal events and 1114 background events
  are left, which result in a sensitivity on the production cross section of 38\%.
  For this specific channel, CLIC's ability to veto events with electrons scattered at very 
  low angles  significantly reduces background contamination, resulting in a sensitivity of 25\%.
\end{description}
A precise measurement of the observables $\sigma_H \times BR(H\to f)$ reflects into a precise determination of the Higgs couplings to the Standard Model bosons and fermions.
The precision achievable on the couplings is usually estimated with a global fit to the cross sections within the so-called $\kappa$-framework\footnote{The parameters $\kappa_i$ are  coupling modifiers, defined as the ratio between the measured and the Standard Model values.}~\cite{k-frame}.
The studies carried out so far at the muon collider do not cover all the most relevant decay modes of the Higgs boson, mainly due to a lack of personnel; therefore, the coupling fit is not yet meaningful.
Nevertheless, this exercise was extremely useful for benchmarking parametric studies that can be used to estimate the sensitivity on the Higgs couplings at a 3 TeV muon collider.

In fact, the parametric study in~\cite{patrick1} presents results for the Higgs boson production cross sections consistent with those for the available channels with the detailed detector simulation. 
It is worth noting that the same study shows that the identification of the $WW$- or $Z^0Z^0$-fusion processes, as discussed in \textbf{Section \ref{sec:muons10tev}}, could significantly improve the coupling sensitivity. 
The estimated precision on the couplings is reported in \textbf{Figure~\ref{fig:Higgs}\textit{d}} in the $\kappa$-framework for a muon collider of 3 TeV 
(1 ab$^{-1}$) and 10 TeV (10 ab$^{-1}$) independently and in combination with the HL-LHC estimates, in the
assumption that there are no BSM decays of the Higgs boson.
While the HL-LHC results are complementary to those of a 3 TeV muon collider,
in the 10 TeV scenario the combination is definitely dominated by the muon collider precision,
with the only exception of the top quark coupling.
This conclusion applies in the perspective of a muon collider directly following the HL-LHC. In the future scenario with an intermediate-energy $e^+e^-$ Higgs factory after the HL-LHC, advocated by both the European Strategy for Particle Physics and the US Snowmass 2021 process, the projected uncertainties in the Higgs boson couplings and decay width are summarized in~\cite{peskin}.

%
%
\subsection{The Higgs boson self-couplings and the Higgs potential}
One of the strengths of the muon collider physics program is the possibility to determine the Higgs field potential.  In the Standard Model~\cite{PDG}, its configuration is parameterized in terms of the vacuum expectation value of the Higgs field
$v=1/\sqrt{\sqrt{2}\,G_F}=246$ GeV, with $G_F$ the Fermi constant, the strength of the Higgs self-coupling $\lambda$, and the Higgs boson mass $m_H = \sqrt{2 \lambda}\, v$. 
Processes in which a virtual Higgs boson produces two or three Higgs bosons, governed by the so-called trilinear and quartic Higgs self-couplings, are of the utmost
important to test whether the Higgs potential is consistent with that predicted by the Standard Model.
To be as general as possible and to account for deviations from the Standard Model, the Higgs potential is formulated with two different values for trilinear ($\lambda_3$) and quartic ($\lambda_4$) self-couplings:
\begin{equation}
    V(h)=\frac{1}{2} m_H^2 h^2 + \lambda_3 vh^3+ \frac{1}{4} \lambda_4 h^4\ ,
\end{equation}  
where $\lambda_3=\lambda_4=\lambda=m_h^2/2v^2$ in the Standard Model and the scalar field $h$ is an expansion around $v/\sqrt{2}$, describing a physical Higgs boson.

As opposed to the case of the Higgs boson couplings to fermions and bosons, high energy is essential for a precise measurement of the Higgs boson self-couplings. 
The gateway to the trilinear self-coupling is the double Higgs production cross section, which is very low at sub-TeV energies.
However, the vector-boson fusion process has a cross section that increases with energy, so measuring the $HH$ production at 3 TeV and even 10 TeV has a great advantage. 
Its measurement was studied with the detailed detector simulation in the hadronic final state $HH\to b\bar{b}b\bar{b}$
for a 3 TeV collider with an integrated luminosity of 1 ab$^{-1}$~\cite{buonincontri}.
Events are selected with at least four reconstructed jets having $p_T^\textsuperscript{jet} > 20$ GeV.
The requirement of $b$-tagging suppresses most of the backgrounds from light quarks.
A final number of 77 signal events and 1422 background events is expected,  which leads to an estimated sensitivity on the production cross section of 33\%.
Including additional final states and assuming an 80\% polarization of the electron beam, CLIC quotes a precision of 22\% with 2 ab$^{-1}$ of data.
From this study, a preliminary estimate in the order of 20-30\% is derived on the sensitivity on the trilinear self-coupling~\cite{sestini}.
It will be possible to determine the final sensitivity with the detailed detector simulation once all the relevant final states are included and the reconstruction algorithms, particularly the $b$-jet tagging, are fully optimized

As the detector design for $\sqrt{s}=10$ TeV is not yet complete, a study was conducted using a parametric description of the physics objects.
In~\cite{tao-H}, the sensitivity at 3 TeV center-of-mass energy has been evaluated with selections that mimic those applied in the analysis with the detailed simulation. The uncertainty on the trilinear self-coupling is $25\%$.
The  exercise is repeated at 10 TeV center-of-mass energy finding an uncertainty of $5.6\%$ assuming 10 ab$^{-1}$ of data.

The muon collider offers a unique opportunity to measure the quartic self-coupling. A parametric study indicates that the quartic self-coupling can be probed to an accuracy of tens of percent with 20 ab$^{-1}$ of data~\cite{chiesa}. Here no background has been taken into account. 
The reconstruction of events with 6 jets, some of them in the forward region, will require dedicated algorithms. 
Additionally, generating the background with a 6 $b$-jet final state is currently not feasible due to the high computational time required.
%
%
\section{THE EXPERIMENT AT THE 10 TEV FRONTIER}
\label{sec:physics-10TeV}
The discovery potential of the multi-TeV muon collider can be exploited by looking for a large variety of signatures \cite{muc_epjc}. The possibility of having collisions at $\sqrt{s}=10$ TeV and beyond enables the search for new particles at an energy scale not accessible by other proposed colliders.
We could consider two different and distinct physics goals for the $\sqrt{s}=10$ TeV muon collider:
\begin{itemize}
\item[-] \textbf{New particles of GeV to $\bm{\sim}$1 TeV mass}: although this mass range is already accessible at the LHC energy, the observation of these particles may have been difficult for several reasons. Going to $\sqrt{s}=10$ TeV muon collisions could enhance the cross-section of such exotic signatures, and the clean environment of lepton collisions could ease their observation. This class includes dark matter particles and other color singlet particles which decay, completely or partially, to invisible final states, as well as axion-like particles. Higgs boson physics could be also included in this category.
\item[-] \textbf{New particles up to 10 TeV mass}: The muon collider at $\sqrt{s}=10$ TeV could be a unique opportunity for measuring new particles from $\sim$1 TeV to 10 TeV mass. In fact, the direct production of such particles is kinematically forbidden at the LHC. The full energy and momentum of such particles could be directly observed in the muon collider detector.
\end{itemize}
From these considerations, it is clear that the capability to measure low-momentum signatures should not be overlooked in the detector design. As an example, scenarios with Higgs composite models or scenarios where dark matter is a particle charged under the electroweak interaction could be probed at the muon collider in a wide range of final state momenta. As will be discussed in this Section, the Standard Model Higgs physics can be considered low momentum physics at $\sqrt{s}=10$ TeV, as well as various other Standard Model measurements. 
The muon collider potential will be fully unleashed with very high energy collisions, and the detector design should be robust for all these kinds of signatures.
%

\subsection{Quest for new physics}
As previously noted, muon collisions at $\sqrt{s}=10$ TeV or higher offer an opportunity to investigate new physics scenarios at the energy frontier and to test several beyond Standard Model hypotheses.
An experimental access to the signatures of many of these scenarios is closely tied to specific capabilities of the experimental apparatus.
Below, such a connection is illustrated for some representative physics cases.

An overview of dark matter candidates that can be searched for at a $\sqrt{s}=10$ muon collider can be found in~\cite{wimp} and has been further developed in \cite{franceschini1, franceschini2, franceschini3}. Within the framework of weakly interacting massive particles, these studies demonstrate that an experiment at a muon collider has the potential to explore models where new electroweak particles were created in the early universe. The analyses are based on inclusive experimental signatures such as mono-photon, mono-muon, dimuon and mono-$W$ events. In particular, it is shown in~\cite{franceschini3} that a $\sqrt{s}=30$ TeV muon collider could exclude the entire coupling space allowed for weakly interacting particles.
A study performed with a detailed detector simulation of mono-photon signatures for searches related to dark photons and axion-like particles~\cite{monophoton} demonstrates the critical importance of detecting objects such as photons with energies in the TeV range. This is crucial for advancing the understanding of new physics, in particular dark matter candidates.

New physics scenarios could manifest themselves with unconventional experimental signatures including, but not limited to, long-lived particles, disappearing tracks, and emerging jets. 
Disappearing tracks refer to charged long-lived particles that travel a few centimeters in the detector before decaying into undetectable particles, resulting in track stubs that do not traverse the entire tracker. The spurious hits produced by the beam-induced background in the tracker can potentially compromise the identification of this type of signature.
The study presented in~\cite{disappearing} demonstrates that the BIB can be effectively mitigated with an appropriate detector configuration and analysis strategy. 
This study has been performed using the detector concept presented in \textbf{Section \ref{sec:detector}}. Further optimization aimed specifically at the 10 TeV center-of-mass energy configuration could improve the results.

The physics possibilities at a $\sqrt{s}\geq10$ TeV muon collider extend far beyond the Higgs sector presented in \textbf{Section~\ref{sec:higgs}} and exotic signatures.
The fact that vector boson collisions dominate the inclusive $\mu^+\mu^-$ interactions at these energies leads to unprecedentedly precise measurements of electroweak processes due to the high production cross section and the relatively clean environment.
This provides a means of indirectly probing new physics by identifying any deviations from the SM predictions.
Measurements of vector boson scattering, the anomalous muon magnetic moment, $b$-hadron decays, and other possibilities can be explored in this context~\cite{muc_epjc}. 
Further studies based on detailed detector simulations
will be necessary in the future to determine the optimal detector configuration and performance for this types of measurements.
%
%
\subsection{Physics and detector requirements for an experiment at $\sqrt{s}=10$ TeV}
\label{sec:det10tev}
\begin{figure}[t]
\centering
\begin{minipage}{0.495\textwidth}
       \hspace{0.7cm}\textbf{\textsf{a}}\vspace{-0.3cm}

        \centering
        \includegraphics[width=1.0\textwidth]{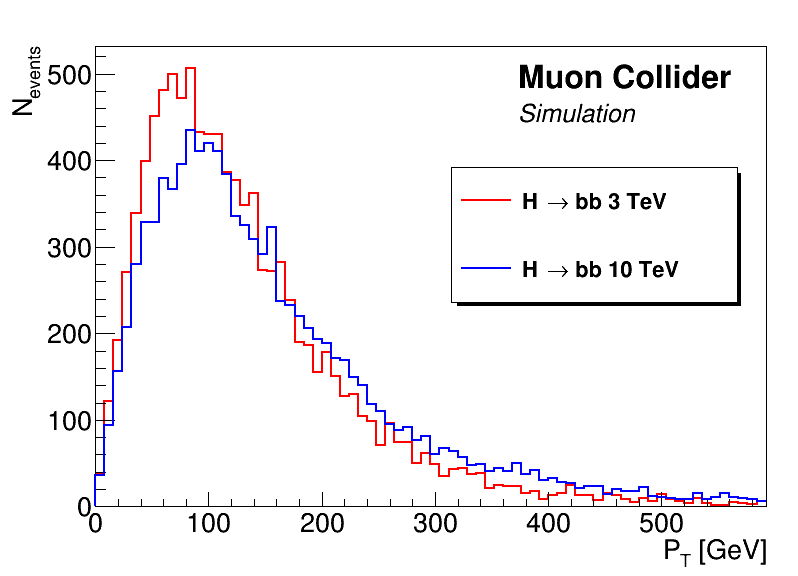}
    \end{minipage} \hfill
    \begin{minipage}{0.495\textwidth}
       \hspace{0.7cm}\textbf{\textsf{b}}\vspace{-0.3cm}
       
        \centering
        \includegraphics[width=1.0\textwidth]{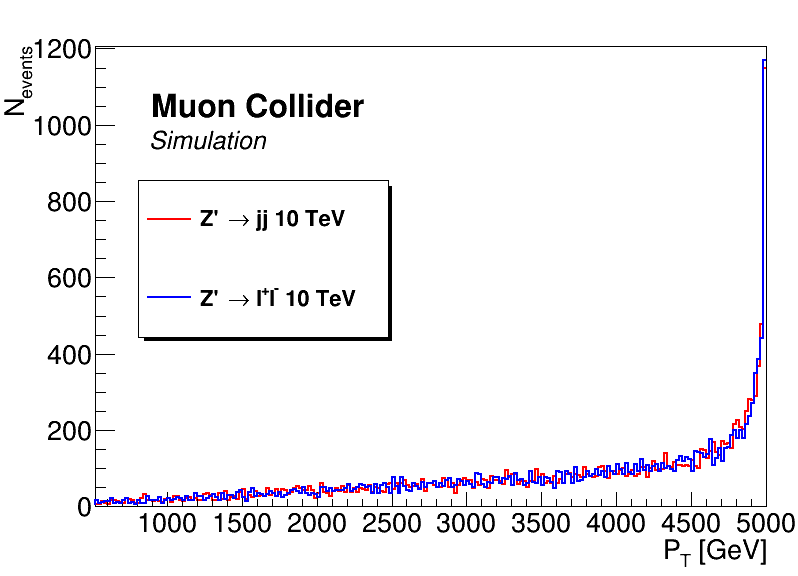}
    \end{minipage}
\begin{minipage}{0.495\textwidth}
       \hspace{0.9cm}\textbf{\textsf{c}}\vspace{-0.2cm}

        \centering
        \includegraphics[width=1.0\textwidth]{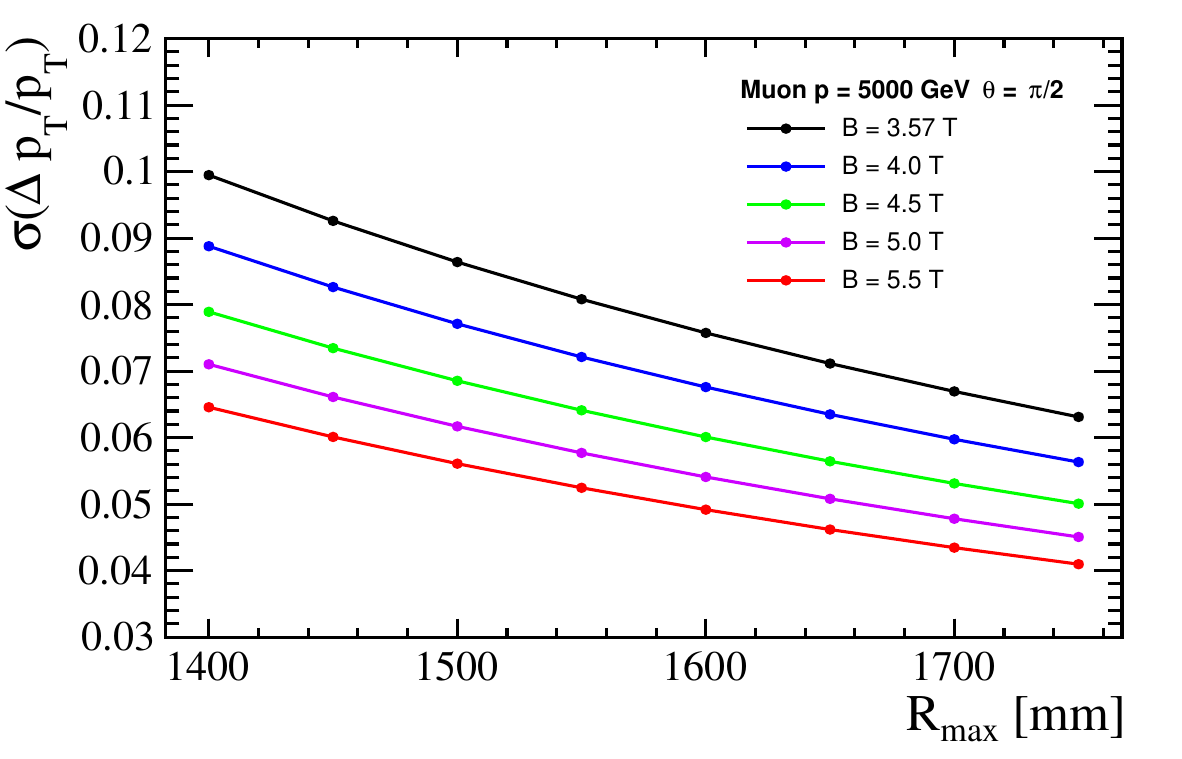}
    \end{minipage} \hfill
    \begin{minipage}{0.495\textwidth}
       \hspace{0.9cm}\textbf{\textsf{d}}\vspace{-0.2cm}
       
        \centering
        \includegraphics[width=1.0\textwidth]{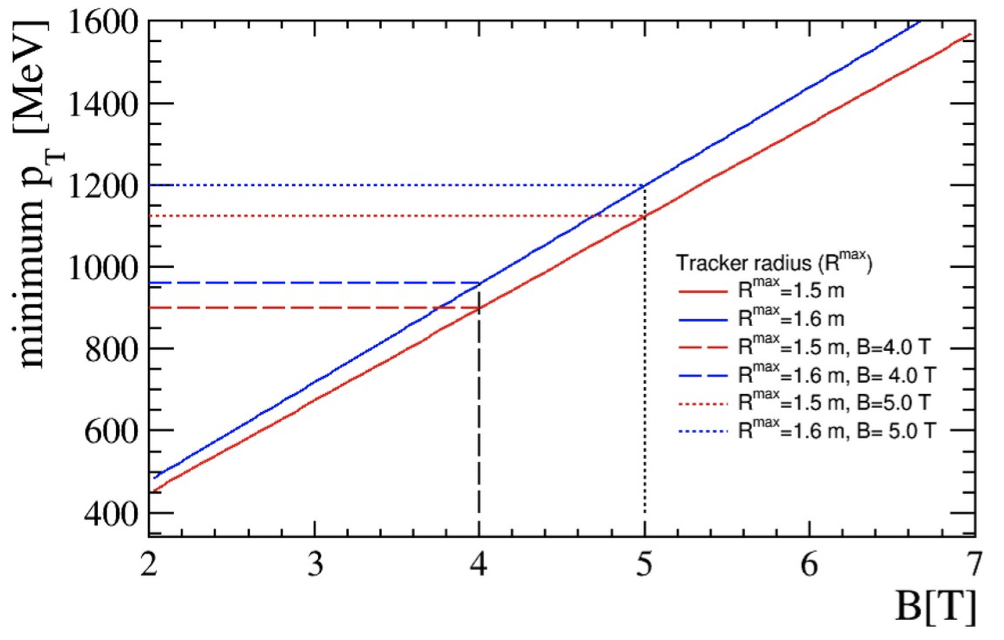}
    \end{minipage}
\caption{\textit{(a)} $b$-quark $p_{\mathrm{T}}$ distributions from the $H \rightarrow b \bar{b}$ decays at $\sqrt{s} = 3$ and 10 TeV, as obtained at generator level. \textit{(b)} Distribution of leptons and jets $p_{\mathrm{T}}$ from the decay of a $Z'$ produced at $\sqrt{s} =10$ TeV with a mass of 9.5 TeV, as obtained at generator level. \textit{(c)} $p_{\mathrm{T}}$ resolution for a muon track with $p_{\mathrm{T}}=5$ TeV and $\theta = \pi/2$ as a function of the tracker radius ($R_{max}$) for different values of the detector magnetic field. \textit{(d)} Minimum $p_{\mathrm{T}}$ required for a charged particle to reach the outermost tracker layer as a function of the magnetic field $B$ for different tracker radius values ($R_{\mathrm{max}}$). Figures taken from \cite{sestinieps}.
\label{fig:10tev}}
\end{figure}

The detector design for an experiment at a muon collider at $\sqrt{s}=10$ TeV is currently under study. In this section, requirements are discussed and guidelines are drawn, but further studies with the detailed simulation are necessary to validate the  detector concept at $\sqrt{s}=10$ TeV. This includes assessing the BIB at detector level, utilizing the $\sqrt{s}=10$ TeV generation described in \textbf{Section \ref{sec:bck}}.

The detector design must aim to provide the best possible efficiency and precision for new physics signatures from low to high momentum, conventional and unconventional. It is also important to ensure the capability to study Higgs physics, as demonstrated with the $\sqrt{s}=$ 3 TeV detector simulation and predicted for $\sqrt{s}=$ 10 TeV (\textbf{Section \ref{sec:higgs}}). 
In the following, as for the case of 3 TeV center-of-mass energy, only one detector concept is discussed. The possibility of having two interaction points may lead to a change in strategy in the future.
 
\textbf{Figure \ref{fig:10tev}\textit{a}} presents the $b$-quark $p_{\mathrm{T}}$ distributions from the $H \rightarrow b \bar{b}$ decays at $\sqrt{s}=3$ and 10 TeV, as obtained at generator level with MadGraph5\_aMC@NLO. Despite the higher center-of-mass energy, it is evident that the distributions are quite similar. Therefore, the considerations made for Higgs physics at 3 TeV also apply to the 10 TeV center-of-mass energy. It can be stated that Higgs physics lives in the low $p_{\mathrm{T}}$ regime even for the $\sqrt{s}=10$ TeV muon collider. Another important feature is that Higgs bosons at $\sqrt{s}=10$ TeV are emitted at smaller polar angles compared to 3 TeV, therefore the forward region of the detector should be designed carefully.

Going from the low energy regime to the high energy scale, \textbf{Figure \ref{fig:10tev}\textit{b}} shows the distributions of leptons and jets $p_{\mathrm{T}}$ from the decays of a $Z^\prime$ produced at $\sqrt{s}=10$ TeV with a mass of 9.5 TeV, as obtained at truth-level with MadGraph5\_aMC@NLO: these distributions exhibit a peak at $p_{\mathrm{T}} \sim 5$ TeV, and a long radiative tail at lower $p_{\mathrm{T}}$. 
A general consideration is that combinatorial and Standard Model backgrounds are expected to have a minor impact at high invariant mass compared to low invariant mass. It is crucial to maintain a good momentum resolution at low $p_{\mathrm{T}}$, while at high $p_{\mathrm{T}}$ a moderate resolution should be sufficient.

Another relevant aspect for the $\sqrt{s}=10$ TeV detector design is that at this high-energy regime the electroweak radiation can be significant. As an example, a muon with a momentum of 3 TeV has a probability of $\sim$30\% to be detected as a $W$ boson~\cite{ew1,ew2,ew3}. 
The possibility of identifying the hadronic decays of the $W$ bosons has not yet been explored, but should be considered in the future studies towards the $\sqrt{s}=10$ TeV detector.

\subsubsection{Tracking system and magnetic field at $\sqrt{s}=10$ TeV}
The handles for improving the track momentum resolution are: 
\begin{itemize}
    \item[-] Tracker transverse dimension.
    \item[-] Number of tracking detector layers. 
    \item[-] Position resolution of the sensor. 
    \item[-] Magnetic field intensity. 
\end{itemize}
To understand the possible tracker configurations, \textbf{Figure \ref{fig:10tev}\textit{c}} shows the $p_{\mathrm{T}}$ resolution as a function of the tracker radius ($R_{\mathrm{max}}$) and magnetic field strength ($B$), as obtained by scaling analytically the detector's performance presented in \textbf{Section \ref{sec:detector}}.
The curves are obtained with the same position resolution assumed for the $\sqrt{s}=3$ TeV detector, as well as the number of tracking layers. The resolution of the hit position depends on the technology of the sensor, and various options have been explored for the muon collider experiment in \cite{technology}.

The choice of the magnetic field is crucial: its magnitude determines the size, construction cost, operational cost, and the technological feasibility of the superconductive solenoid. As a rule of thumb, the energy density of the magnetic field is proportional to $B^2$, and the solenoid cost has approximately the same scaling. Very high magnetic fields require the development of challenging technologies to maintain the mechanical and electrical stability of the superconducting coil, and to achieve uniformity in the field.  For these reasons, it is important to aim for the lowest possible magnetic field strength, as long as it allows the desired physics performance to be achieved.

Increasing the magnetic field intensity enhances the tracking momentum resolution, but there is also a trade-off at low $p_{\mathrm{T}}$: the minimum $p_{\mathrm{T}}$ required for a charged particle to reach the outermost layer of the tracker is proportional to $B$. In \textbf{Figure \ref{fig:10tev}\textit{d}}, the minimum $p_{\mathrm{T}}$ as a function of $B$ for different tracker radius hypotheses is shown. As an example, with $B=5$ T and $R_{\mathrm{max}}= 1.6$ m, the minimum $p_{\mathrm{T}}$ is 1.2 GeV. Below this threshold, a deterioration in efficiency and/or resolution is expected. Preliminary results~\cite{sestinieps} indicate  that going from $B=3.57$ T to $B=5$ T, assuming the detector configuration of \textbf{Section \ref{sec:detector}}, results in  an efficiency loss of about 10$\%$ for reconstructing tracks originating from the $H \rightarrow b \bar{b}$ decay.
Based on the above considerations, a magnetic field strength between 4 and 5 T and a tracker radius larger than that of the $\sqrt{s}=3$ TeV detector could be a good compromise at $\sqrt{s}=10$ TeV. For instance, a radius of $R_{\mathrm{max}}=1.6$ m could be a suitable choice.

\subsubsection{Calorimeter systems at $\sqrt{s}=10$ TeV} 
Showers produced by photons, electrons, and hadrons with energies of the order of TeV need be contained in the calorimeter volume to achieve an adequate resolution on the energy measurement.
For example, a photon with an energy of 1 TeV deposits less than 95\% of its energy in 22 $X_0$ of ECAL \cite{calo_containment}, the depth of the electromagnetic calorimeter described in \textbf{Section~\ref{sec:detector}}. The remaining energy spills into the HCAL. It is evident that the detector at $\sqrt{s}=10$ TeV should have deeper calorimeters compared to the $\sqrt{s}=3$ TeV configuration. Additionally, alternative materials with shorter $X_0$ values should be considered. In fact, it is important to remark that the dimensions of the detector are constrained by the cavern dimension, the layout of the accelerator components, the MDI and the magnet structure. 

Another aspect to consider is the position of the solenoid with respect to the calorimeters. Placing the solenoid in front of the ECAL could shield the calorimeter from the low-energy BIB photons, but could also absorb the photons from $H \rightarrow \gamma \gamma$ with lower energies or the electromagnetic component of $b$-jets from $H \rightarrow b \bar{b}$, degrading the performance in the energy range where a precise resolution is crucial. On the other hand, positioning a high-field solenoid outside HCAL might be costly due to its large dimensions. 
A compromise could be placing the solenoid between ECAL and HCAL. However, defining the final configuration requires detailed simulation studies and the assessment of BIB impact.

Several calorimeter technologies that meet the requirements of a multi-TeV muon collider are currently under study. These technologies are not only being explored through simulations, but are also characterized with experimental tests. For example, an ECAL made of $PbF_2$ crystals has been proposed in~\cite{crilin}, and an evaluation of Micro Pattern Gas detectors for the active layers of HCAL is ongoing~\cite{bari}.

\subsubsection{Muon detection at $\sqrt{s}=10$ TeV} 
\label{sec:muons10tev}
Identifying and measuring muons with momenta in the TeV range present significant challenges. 
Promising technologies are discussed in \cite{picosec}, where gas detectors with excellent time resolution and high rate capability are presented.
In the muon detector, the bending due to the residual magnetic field alone is not sufficient for the precise determination of muon $p_{\mathrm{T}}$. Nevertheless, the muon detector can still be used to tag events with high momentum muons.

Additionally, high-granularity imaging calorimeters~\cite{calice} could play an important role in detecting high-momentum muons since significant radiative losses are expected for these particles with $p_{\mathrm{T}}$ in the TeV range.
Therefore, integrating information from all subsystems, including tracker, calorimeters and muon detectors, will be of paramount importance for muon reconstruction. It is essential to develop  algorithms optimized for the TeV scale.

The detection of forward muons, \textit{i.e} muons that are scattered at very low angles with respect to the beam direction and pass through the nozzles, would be an important feature of the muon collider detector. 
This capability would make it possible to distinguish between the $WW$ and $Z^0Z^0$ fusion processes. The former is accompanied by forward neutrinos, while the latter has forward muons.
At $\sqrt{s}=10$ TeV, the muons from $Z^0Z^0$-fusion process are expected to be even more forward compared to $\sqrt{s}=3$ TeV. Consequently, the detector acceptance may cover just a small fraction of the angular distribution. The feasibility of instrumenting the region close to the beamline to tag these muons is currently under investigation.
%
%
\section{OUTLOOK}
The muon collider facility, initially proposed by MAP, has been revised during the Accelerator R\&D Roadmap~\cite{LDG} and by the IMCC, revealing no fundamental showstoppers. 
Using the same scheme as MAP for muon production, a preliminary design of a machine at a center-of-mass energy of 10 TeV was possible, opening up uncharted territory for lepton collisions. 

For the first time, extensive studies have been conducted on detector performance in the presence of beam-induced background. Currently, the BIB simulation sample generated by MAP is still used, due to the complexity of the generation process. Given the critical impact of BIB on the detector, extensive validations were done by using a different simulation software leading to consistent results. This validation gives confidence and robustness in the predictions of the beam-induced background.
The proposed detector concept for $\sqrt{s}=3$ TeV and the associated physics object reconstruction algorithms have demonstrated to be able to handle the effects of the beam-induced background, although neither the detector nor the algorithms have been optimized yet.

The physics performance assessed at $\sqrt{s}=3$ TeV with the detailed detector simulation, including the effects of the beam-induced background, pertains mainly the Higgs physics. It was imperative to demonstrate that such a detector in such an environment has excellent capabilities to study an existing particle. Other studies are currently in development, but have not yet been carried out due to lack of personnel.

Nonetheless, this has demonstrated that an initial stage of a muon collider operating at a center-of-mass energy of 3 TeV can achieve a performance comparable to that of the extensively studied CLIC detector at the same collision energy.
While this energy is seen as the ultimate limit for $e^+e^-$ colliders due to beamstrhalung radiation, it is considered the first stage for the muon collider.

The results obtained at $\sqrt{s}=3$ TeV were used to benchmark parametric studies of the Higgs processes not included in the studies with the detailed detector simulation. Such studies were carried out at the same center-of-mass energy and at $\sqrt{s}=10$ TeV and beyond. 
The possibility to reach high-energy muon collisions allows to probe new physics by investigating various models including the precise determination of the Higgs potential.

A detector concept for muon collisions at 10 TeV center-of-mass energy is currently in preparation, the requirements were studied as discussed in this review for the first time. At $\sqrt{s}=10$ TeV, the detector shielding structure needs to be designed, currently the nozzles optimized by MAP for $\sqrt{s}=1.5$ TeV are used across all energies. Designing a new nozzle requires expertise that had been lost after MAP shutdown, only recently IMCC recovered it. Preliminary studies indicate that the amount of BIB particles arriving at the detector is not significantly larger than at $\sqrt{s}=3$ TeV, the aim is to reduce the dimension of the tungsten material to enhance the detector acceptance in the forward region. This region is one of the open questions for a detector at the muon collider. In fact, there are physics processes, as for example the Higgs physics, that at high energy, $\sqrt{s} \geq 10$ TeV, are produced with higher cross section in the forward region. It would also be important to have at least the possibility to tag forward muons to distinguish between $WW$ from $Z^0Z^0$ boson fusion production.

The design of the detector at $\sqrt{s}\geq 10$ TeV has demonstrated that the requirements for the technologies needed to achieve the desired performance could be satisfied through the R\&D processes already ongoing for HL-LHC and future colliders. The muon collider community is actively engaged in the \textit{Detector Research and Development collaborations - DRD}~\cite{DRD} in Europe and in the \textit{Coordinating Panel for Advanced Detectors - CPAD}~\cite{CPAD} in the United States. Participation in the development of new detector designs and related test beams for a novel detector concept is of crucial importance to engage and sustain the young community.

On a shorter time scale, the muon collider community may have to design and operate a facility to demonstrate the full muon ionization cooling chain. Experts in accelerator and detector technologies will need to collaborate together to address challenges previously not encountered. 
Moreover, both the detector R\&D and the demonstration facility will play a vital role in actively involving the community in experimental activities, thereby preventing the loss of valuable expertise and knowledge.
%
\section*{DISCLOSURE STATEMENT}
The authors are not aware of any affiliations, memberships, funding, or financial holdings that
might be perceived as affecting the objectivity of this review. 
%
\section*{ACKNOWLEDGMENTS}
We would like to thank the many people who contributed with their hard work, useful discussions, and valuable advice:
Chiara Aim\'e, Paolo Andreetto, Nazar Bartosik, Laura Buonincontri, Daniele Calzolari, Luca Castelli, 
Giacomo Da Molin, Luca Giambastiani, Alessio Gianelle, Sergo Jindariani, Karol Kritza, Anton Lechner, Alessandro Montella, Mark Palmer,
Simone Pagan Griso, Nadia Pastrone, Cristina Riccardi, Ilaria Vai, Davide Zuliani.
We are grateful to the International Muon Collider Collaboration and the US Muon Accelerator Program for their support.
We acknowledge the financial support of the Italian National Institute for Nuclear Physics (INFN), the University of Padua, 
and the European Organization for Nuclear Research (CERN).
This work was supported by the European Union’s Horizon 2020 and Horizon Europe Research and
Innovation programs through the Marie Skłodowska-Curie RISE Grant Agreement No. 101006726, the Research Infrastructures 
INFRADEV Grant Agreement No. 101094300, and the EXCELLENT SCIENCE - Research Infrastructures Research Innovation Grant Agreement No. 101004761.
%

%
\end{document}